\documentclass[12pt, reqno,a4paper]{amsart}
\usepackage[lmargin=3cm,rmargin=3cm,bmargin=3cm,tmargin=2.5cm]{geometry}

\usepackage{incgraph}

\usepackage{tabularx}
\usepackage[utf8]{inputenc}
\usepackage{amsmath}
\usepackage{amssymb}
\usepackage{amsthm}
\usepackage{textcomp}
\usepackage{enumerate}
\usepackage[numbers]{natbib}
\usepackage{array}
\usepackage{bbm}
\usepackage{verbatim}
\usepackage{hyperref}
\usepackage{color} 
\usepackage{graphicx}
\usepackage{booktabs}
\usepackage{upgreek}

\usepackage{datetime}
\usepackage{enumerate}
\usepackage{subcaption}
\usepackage{tikz-cd}
\usepackage{float}
\floatstyle{ruled}
\newfloat{algorithm}{t}{lop}
\floatname{algorithm}{Algorithm}
\usepackage{thmtools,thm-restate}
\usetikzlibrary{shapes,arrows,snakes,backgrounds}
\usepackage{mathrsfs} 
\usepackage{bibentry}
\usepackage{pdfpages}

\hypersetup{
  colorlinks,
  linkcolor={red!50!black},
  citecolor={blue!50!black},
  urlcolor={blue!80!black}
}

\providecommand{\abs}[1]{\lvert#1\rvert}
\providecommand{\norm}[1]{\left\lVert#1\right\rVert}

\newcommand\Cdot{\ensuremath{%
  \mathchoice%
   {\mskip\thinmuskip\lower0.2ex\hbox{\scalebox{1.5}{$\cdot$}}\mskip\thinmuskip}}%
   {\mskip\thinmuskip\lower0.2ex\hbox{\scalebox{1.5}{$\cdot$}}\mskip\thinmuskip}%
   {\lower0.3ex\hbox{\scalebox{1.2}{$\cdot$}}}%
   {\lower0.3ex\hbox{\scalebox{1.2}{$\cdot$}}}%
}

\newcommand{\mulp}{I}
\newcommand{\wmarg}{\dot{w}}

\newcommand{\defeq}{\mathrel{\mathop:}=}

\newcommand{\X}{\mathsf{X}}
\newcommand{\T}{\mathsf{T}}
\newcommand{\pr}{\mathrm{pr}}

\newcommand{\Y}{\mathsf{Y}}

\newcommand{\cE}{\mathcal{E}}

\newcommand{\uarg}{\,\cdot\,}
\newcommand{\ud}{\mathrm{d}}

\newcommand{\R}{\mathbb{R}}
\newcommand{\N}{\mathbb{N}}
\renewcommand{\P}{\mathbb{P}}
\newcommand{\E}{\mathbb{E}}

\newcommand{\eps}{\epsilon}

\newcommand{\charfun}[1]{\mathbf{1}\big\{#1\big\}}

\newcommand{\var}{\mathrm{var}}

\newcommand{\inner}[2]{\left\langle #1, #2 \right\rangle}

\newcommand{\const}[1]{c_{#1}}
\newcommand{\lowc}{\underline{c}}
\newcommand{\highc}{\bar{c}}

\theoremstyle{remark}

\theoremstyle{definition}

\newtheorem{main-lemma}{Lemma}
\newtheorem{main-theorem}{Theorem}

\renewcommand{\cE}{\mathcal{E}}

\newcommand{\bX}{{\mathbf{X}}}
\newcommand{\bx}{{\mathbf{x}}}

\begin{document}
\nobibliography*

\incgraph[documentpaper,
  overlay={}]
         [width=\paperwidth,height=\paperheight]{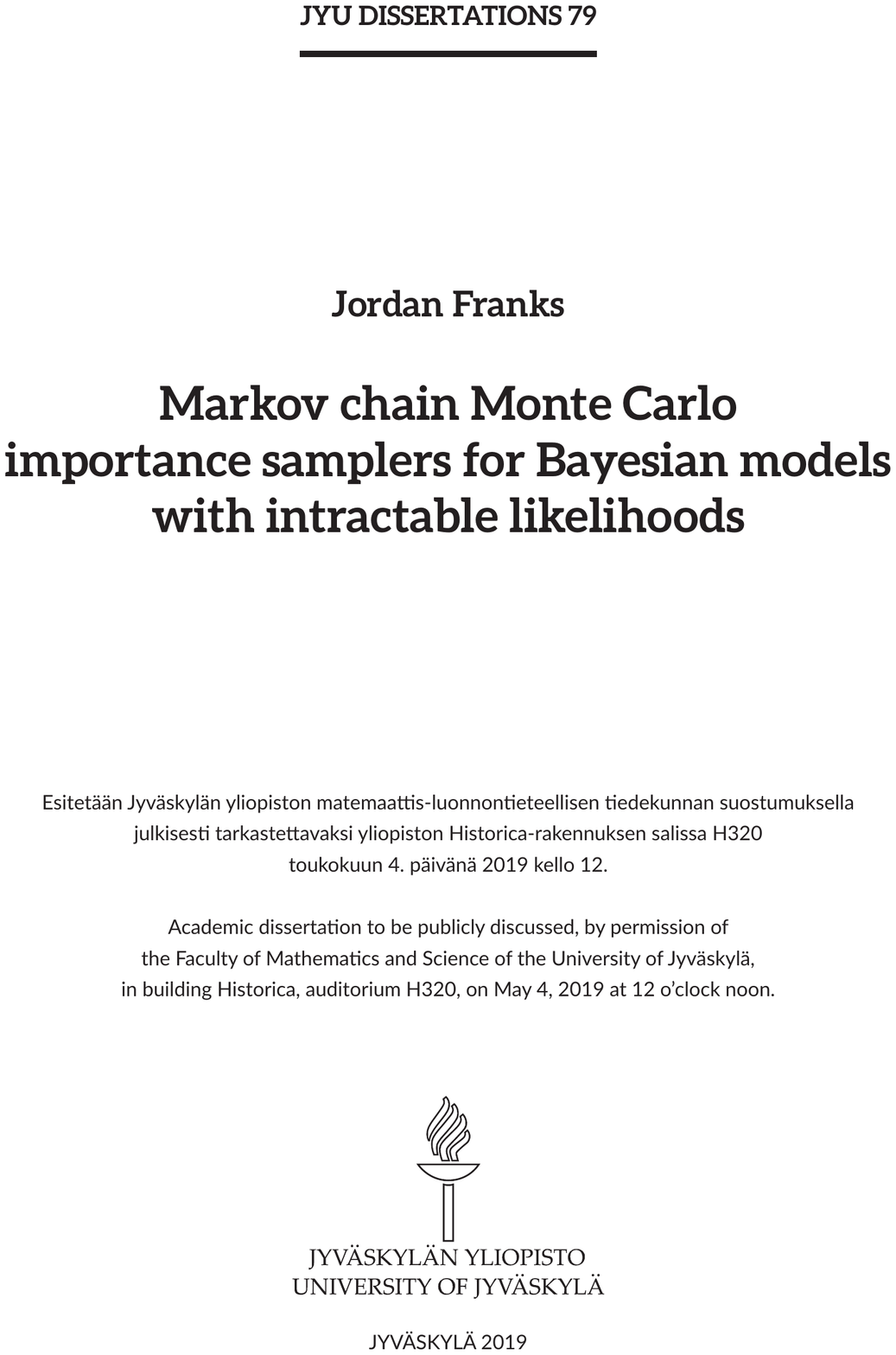}
         \newpage
         \incgraph[documentpaper,
  overlay={}]
  [width=\paperwidth,height=\paperheight]{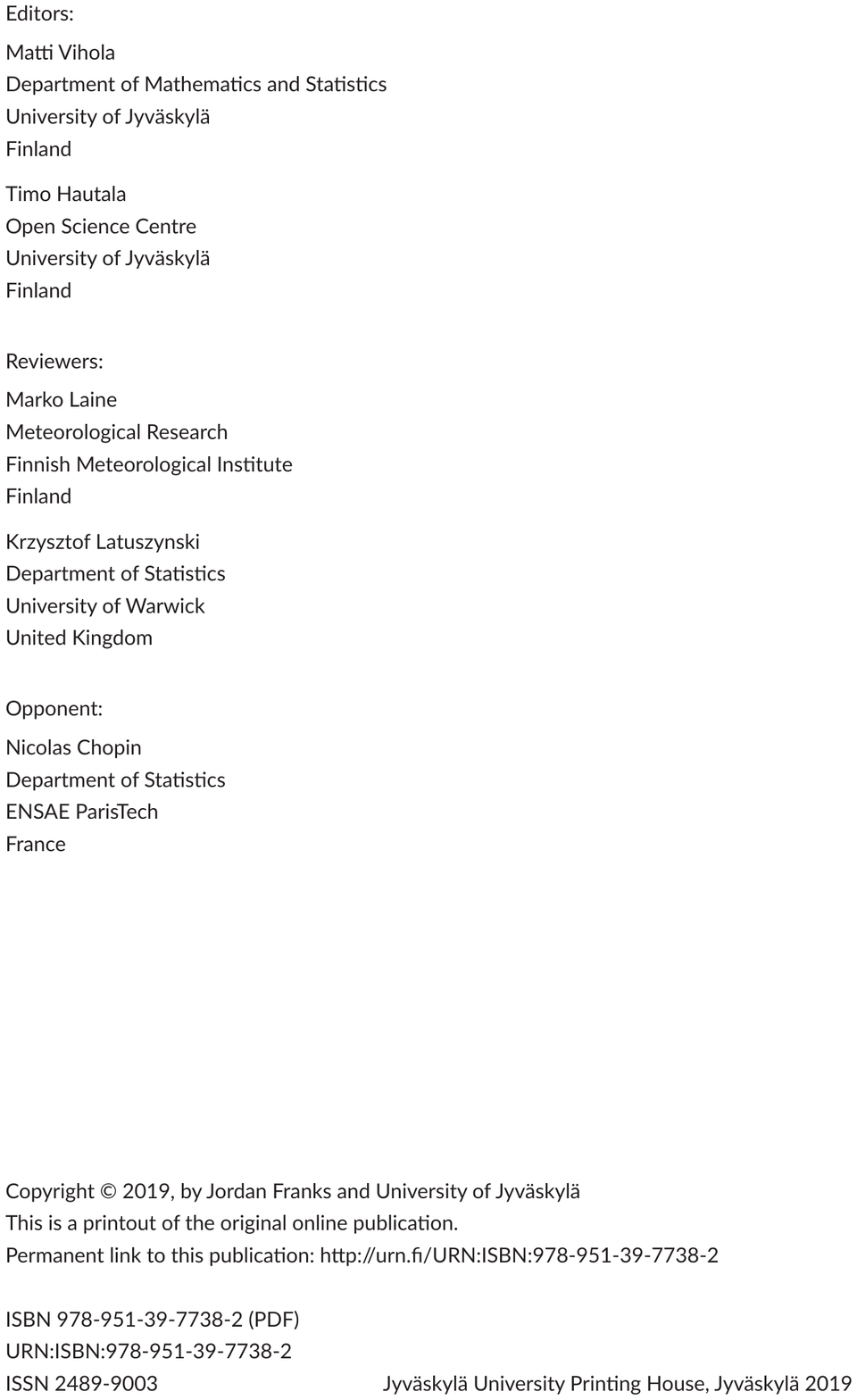}

 \title[MCMC importance samplers]{}

\author[Jordan Franks]{}

\maketitle
\thispagestyle{empty}
\pagenumbering{roman}




\vspace*{10pt}
\section*{Abstract}
Markov chain Monte Carlo (MCMC) is an approach to parameter inference in Bayesian models
that is based on computing ergodic averages formed from a Markov chain targeting the Bayesian posterior probability.
  We consider the efficient use of an approximation within the Markov chain, with subsequent importance sampling (IS) correction of the Markov chain inexact output, leading to asymptotically exact inference.
We detail convergence and central limit theorems for the resulting MCMC-IS estimators.
We also consider the case where the approximate Markov chain is pseudo-marginal, requiring unbiased estimators for its approximate marginal target.  Convergence results with asymptotic variance formulae are shown for this case, and for the case where
the IS weights based on unbiased estimators are only calculated for distinct output samples of the so-called `jump' chain, which, with a suitable reweighting, allows for improved efficiency.  As the IS type weights may assume negative values, extended classes of unbiased estimators may be used for the IS type correction, such as those obtained from randomised multilevel Monte Carlo.  Using Euler approximations and coupling of particle filters, we apply the resulting estimator using randomised weights to the problem of parameter inference for partially observed  It\^{o} diffusions.
Convergence of the estimator is verified to hold under regularity assumptions which do not require that the diffusion can be simulated exactly.  
In the context of approximate Bayesian computation (ABC), we suggest an adaptive MCMC approach to deal with the selection of a suitably large tolerance, with IS correction possible to finer tolerance, and with provided approximate confidence intervals.  
A prominent question is the efficiency of MCMC-IS compared to standard direct MCMC, such as  pseudo-marginal, delayed acceptance, and ABC-MCMC. 
We provide a comparison criterion which generalises the covariance ordering to the IS setting.  We give an asymptotic variance bound relating MCMC-IS with the latter chains, as long as the ratio of the true likelihood to the approximate likelihood is bounded.  We also perform various experiments in the state space model and ABC context, which confirm the validity and competitiveness of the suggested MCMC-IS estimators in practice. 

\vspace*{\fill}
\newpage

\normalsize
\tableofcontents
\normalsize
\section*{List of symbols}
\begin{table}[H]
  \begin{tabular}{l c l}
    Symbol & Page & Meaning \\
    \midrule
    $(\Y,\mathcal{Y}, \mathscr{P})$ & \pageref{sec:intro-likelihoods} & statistical model \\
    $\mathscr{P}$ & \pageref{eq:family} & family of probability distributions\\
    $p^{(\theta)}$ & \pageref{eq:family} & data (probability) distribution \\
    $L(\theta)$ & \pageref{eq:likelihood} & likelihood of parameter $\theta$ \\
    $\mathscr{P}_\ell$& \pageref{eq:ell-approx-family} & $\ell$-approximate family of data (probability) distributions\\
    $\mathscr{P}_\infty$ & \pageref{eq:ideal-family} & ideal family of data (probability) distributions\\
$(M_p, G_p)$ & \pageref{eq:fk-potential} & Feynman-Kac model with transition $M_p$ and potential $G_p$\\
    $p^{(\theta,\ell)}$ & \pageref{eq:ell-approx-family} & data (probability) distribution in $\mathscr{P}_\ell$\\
    &  \pageref{eq:latent-smoother}& $\ell$-smoother for Feynman-Kac model $(M_p^{(\theta,\ell)},G_p^{(\theta)})$\\
    $p_u^{(\theta,\ell)}$ &\pageref{eq:ell-un-smoother} & unnormalised $\ell$-smoother on latent states for model $(M_p^{(\theta,\ell)},G_p^{(\theta)})$\\
    $\pi^{(\ell)}$ & \pageref{eq:ell-posterior} & joint $\ell$-posterior probability on parameters and latent states\\
    $\pi_m^{(\ell)}$ & \pageref{eq:marginal-posterior} & marginal $\ell$-posterior probability on parameters\\
    $p_u$ & \pageref{eq:un-smoother} & unnormalised smoother for model $(M_p,G_p)$\\
        $\hat{p}_u$ & \pageref{eq:p-hat} & unbiased estimator for $p_u$ for model $(M_p,G_p)$\\
    $\hat{L}(\theta)$ & \pageref{eq:likelihood-estimator} & unbiased estimator for $L(\theta)$ \\
    $\sigma_K^2$ & \pageref{eq:clt} & asymptotic variance for Markov kernel $K$\\
        $\mathcal{E}_K$ & \pageref{eq:dirichlet-form} & Dirichlet form for Markov kernel $K$ \\
    $\Pi^{(0)}$ & \pageref{sec:comparison-peskun-type} & stationary probability of $0$-model Markov chain\\
        $\Pi^{(\infty)}$ & \pageref{sec:comparison-peskun-type} & stationary probability of $\infty$-model Markov chain\\
    $w$ & \pageref{eq:importance-sampling-weight} & importance sampling weight from $\Pi^{(0)}$ to $\Pi^{(\infty)}$\\
    $\sigma_{IS}^2$ & \pageref{eq:is-asvar} & asymptotic variance of MCMC-IS estimator \\
    $\Delta^{(\theta,\ell)}$ & \pageref{eq:delta-pf-estimator} & delta particle filter unbiased estimator\\
        $(p_\ell)_{\ell\in \N}$ & \pageref{sec:diffusions-debiasing} & probability mass function on $\ell=1,2,3,\ldots$\\
    $    \tilde{\Delta}^{(\theta,\ell)}(\phi)$ & \pageref{eq:tilde-delta} & unbiased estimator for $p_u^{(\theta,\ell)}(\phi)$\\
    $\mathscr{C}(m)$ & \pageref{eq:total-computational-cost} & total computational cost for $m$ iterations\\
    $\mathscr{M}(\kappa)$& \pageref{eq:realised-length} & realised length of the chain with budget $\kappa$ \\
    $p^{(\theta,1/ \epsilon)}$ & \pageref{eq:abc-data-dist} & data (probability) distribution, with ABC tolerance $\epsilon$ \\&&(or precision $1/\epsilon$)\\
    $L^{(1/\epsilon)}$ & \pageref{eq:abc-likelihood} & approximate Bayesian computation (ABC)  likelihood \\&& with ABC tolerance $\epsilon$ (or precision $1/\epsilon$)
  \end{tabular}
  \end{table}
\normalsize
\newpage
\section*{Foreword}

The modern reliance on probability theory to model the universe and various aspects of life reveals on the one hand our tremendous lack of knowledge and ability to understand and hence predict the workings of the universe with Newtonian precision.  On the other hand, the success of probability theory reveals the hidden order of the universe, as well as the significant deductive reasoning capacities of humankind, where from the disorder of incomplete knowledge arises the order of probabilistic laws.
Statistics allows us to ascertain to what extent our deductive reasoning is justified by real observation.  Statistics acts as the intermediary, allowing dialogue to proceed between our perceived knowledge of the (mechanistic and probabilistic) laws of the universe and of the universe as she presents herself to us in actual fact. %

Of utmost importance for the development of statistics has been the increasing computational ability in the computer age \citep[cf.][]{efron-hastie}.  As the speed of computers increases, so does the potential complexity of problems increase which statistical methods can handle with precision.  Considerable interest therefore lies in the development of computational methods which are efficient and able to perform the demanding computational tasks of modern statistics.  It is the scientific and humanistic hope for this thesis, that the work will serve to the advancement of human knowledge, and that it will be solely useful to the commendable pursuits of humankind.

As the fields of probability and statistics are intellectually challenging, any small progress in this field is dependent upon a stable, friendly, and stimulating working and living environment.
First of all, I would like to thank Dr.~Matti Vihola, for being a wonderful adviser, scientist, and person.  
In the beginning, I knew very little about computational statistics and Monte Carlo methods, but due greatly to Matti's tremendous help and patience, my knowledge and skills in mathematics and statistics has grown considerably.  This thesis would not have been possible without his help.  The enclosed introductory text has also benefited greatly from his insightful remarks.
As his first sole doctoral student, I have one of the early claims to be able to call him mathematico-statistical father.

As for a stable, friendly, and stimulating working and living environment, I would like to thank the University of Jyväskylä and its employees, for being able to pursue my doctoral studies here.  The last three years have been enjoyable as a place to work, study, and  live.
Financial support is gratefully acknowledged from the Academy of Finland (`Exact approximate Monte Carlo methods for complex Bayesian inference,' grants 284513 and 312605, PI: Dr.~Matti Vihola).

Sincere thanks to the reviewers, Prof.~Marko Laine (Finnish Meteorological Institute) and Prof.~Krzysztof Łatuszy{\'n}ski (University of Warwick).

There are many other individual persons whom I should thank for being a help these last few years. 
As I drew up an account of all the people whom I would like to thank, it became ever-expanding, touching every aspect and time of my life.  I simply could not do proper justice to those who have helped me, and I would run the danger of leaving somebody unintentionally out.  I therefore would simply like to thank the many precious people who have been a positive impact on me, without going into all the details here.  They and there deeds are simply too many to be entrusted to these few pages.

I think the saying is true, and hope it is true:  when someone has stayed somewhere long enough, the place becomes forever a part of the person.  I wish to thank the many people in Jyväskylä whom I have enjoyed getting to know during these last few years.  Language has not always been an insurmountable barrier.
I will miss you, and I will miss Jyväskylä---the snow, the sauna, the summer, the lakes, the festivals, the food, the people---all of which make Finland a special place to live.

I mention regards to the researchers everywhere with whom I have had the privilege to meet.
Statistics, like other scientific disciplines such as pure mathematics, involves many devotees interested in a common subject with undesirable distractions kept to a minimum.
When immersed in a scientific subject, where validity is judged by logic and observation rather than might or necessity, when one is able to escape the day-to-day absorption of the human condition, then one is able to view the world from a new perspective.
One sees like the astronaut, for whom, after seeing the Earth as the single terrestrial mass, life will never be the same.

\bigskip
\bigskip
\bigskip
\bigskip
\begin{flushleft}
  Jordan Franks\\
  Jyv\"askyl\"a, Finland\\
April 4, 2019
\end{flushleft}

\newpage
\section*{List of included articles}\label{sec:list}
The thesis consists of an introductory text (Sections \ref{sec:intro}-\ref{sec:summary}) and the following articles listed in chronological order of preprint appearance.
\let\oldaddcontentsline\addcontentsline
\renewcommand{\addcontentsline}[3]{}

\let\addcontentsline\oldaddcontentsline
\bigskip

In the introduction, the above articles are referred to by letters \cite{viholaHF}, \cite{franks-vihola}, \cite{franksJLV}, and \cite{vihola-franks}, whereas other article references cited in the introduction shall be referred to by numbers [1], [2], [3], etc. 

The author of this dissertation has actively contributed to the research of the joint articles \cite{viholaHF}, \cite{franks-vihola}, \cite{franksJLV} and \cite{vihola-franks},
and has served as the corresponding author for articles \cite{franks-vihola} and \cite{franksJLV}.
In particular, the author contributed some theoretical results for augmented Markov chains for \cite{viholaHF}, and assisted during the research and preparation of the article.  Article \cite{franks-vihola} is mostly the author's own work, but benefited greatly from the second author's input, for example, regarding the topic of the paper and some technical details.  The author was responsible for the efficiency considerations of multilevel Monte Carlo in \cite{franksJLV}, and for final preparation of the article.  The author was responsible for the analysis of the adaptive Markov chain Monte Carlo algorithm in \cite{vihola-franks}.

\newpage


\pagenumbering{arabic}

\section{Introduction}\label{sec:intro} 
Bayesian inference often requires the use of computational simulation methods known as Monte Carlo methods.
Monte Carlo methods use probabilistic sampling mechanisms and ensemble averages to estimate expectations, such as those taken with respect to the posterior probability of a Bayesian model.  Therefore, in practice on a computer, Monte Carlo methods can be computationally intensive.

A further inferential challenge arises when the likelihood function of the Bayesian model is intractable.  In some important settings, it is possible to obtain an unbiased estimator for the likelihood.
One such setting is the state space model, where sequential Monte Carlo supplies the unbiased estimator.  In settings where unbiased estimators are not possible, approximate Bayesian computation (ABC) may be used,
assuming forward generation of the  model is possible and a choice of tolerance size has been made.  Though only an approximation to the original Bayesian model, the ABC model comes equipped with a straightforward unbiased estimator for its ABC likelihood.

In these two example settings, a Markov chain can be run, allowing for Markov dependence in the samples, as well as the use of the unbiased estimators for the (ABC) likelihood as part of a pseudo-marginal algorithm.
As a result, the samples of the Markov chain are drawn asymptotically from the (ABC) posterior, and inference is based on averaging the samples obtained.  This computational approach to Bayesian inference is known as Markov chain Monte Carlo (MCMC).

This thesis is  concerned with a slightly different approach, namely, where the Markov chain targets an approximate marginal of the (ABC) posterior.
The subsequent importance sampling (IS) type correction is performed by a reweighting of the inexact sample output, using the unbiased estimators, which yields asymptotically exact inference for the (ABC) posterior.
The use of an approximation for the Markov chain target can be computationally efficient, as can be the parallel calculation of the IS weights on a parallel computer.  
Some of the resulting MCMC-IS estimators are well-known, but in practice have been used only rarely, in comparison to direct MCMC.  In addition, the MCMC-IS approach is shown to offer additional flexibility compared to direct MCMC.

The rest of this Section \ref{sec:intro} is laid out as follows.
We present some important notions, such as that of a statistical model, likelihood function, and Bayesian model.
We briefly describe the general goal of (likelihood-based) parameter inference in statistics,  as well as some of the challenges of computation which the thesis seeks to address, specifically inference aided by use of an approximation.  Section \ref{sec:intro-overview} concludes with an outline of the remainder of the text.


\subsection{Likelihoods}\label{sec:intro-likelihoods}
A \emph{statistical model} $(\Y,\mathcal{Y}, \mathscr{P})$ is composed of an \emph{observational space} $\Y$, together with its $\sigma$-algebra of subsets $\mathcal{Y}$, and a set $\mathscr{P}$ of probability distributions on $\Y$ \cite[cf.][]{georgii}.  We assume the family $\mathscr{P}$ is parametrised by a vector of \emph{model parameters} $\theta\in\T$, with $\T\subset \R^{n_\theta}$ for some $n_\theta\ge 1$.  That is,
\begin{equation}\label{eq:family}
  \nonumber
\mathscr{P} = \{ p^{(\theta)} \}_{\theta\in\T},
\end{equation}
where $p^{(\theta)}(\ud y)$ is a probability on $\Y$, sometimes called the \emph{data distribution}.  The probability $p^{(\theta)}$ corresponds to a modeling of the dependency relationship of the observation $y$, considered as a random variable, on the model parameter $\theta$.

We assume for simplicity in this introduction that $p^{(\theta)}(\ud y)$  has a density, also denoted $p^{(\theta)}(y)$, with respect to a $\sigma$-finite reference measure on $\Y$.
Fixing the observation $y\in\Y$, we define the function
\begin{equation}\label{eq:likelihood}\nonumber
L(\theta)\defeq p^{(\theta)}(y),
\end{equation}
which is known as the  \emph{likelihood}.
 One type of likelihood-based inference for $\theta$ is to answer which values of $\theta$  maximise $L(\theta)$.  This method of inference is known as \emph{maximum likelihood estimation} (MLE) in a statistical model with observation \citep[cf.][]{cappeMR}.  In other words, MLE seeks to answer, which  probability distribution on $\Y$ in $\mathscr{P}$ would most readily give rise to the observation.  


\subsection{Bayesian inference}\label{sec:intro-bayesian}
In practice, MLE is highly dependent on the candidate set $\mathscr{P}$ of probabilities to consider.  The set $\mathscr{P}$ could be parametrised by arbitrarily high dimensions of parameters, and is the result of the statistician's modeling of the dependence of the observation $y$ on the model parameter $\theta$.  Going further, in light of this arbitrary construction of the set $\mathscr{P}$, the statistician is arguably\footnote{The \emph{frequentist} approach differs from the \emph{Bayesian} approach considered here  \citep[cf.][]{efron-hastie}.}  not out of bounds to specify which $\theta$ values are to be considered more probable and with more weight,  based on prior knowledge or hypothesis.

This specification, for a statistical model with known observation, leads to the \emph{Bayesian model}  \citep[cf.][]{gelmanCSR}.  The \emph{Bayesian model} consists of an assignment of a \emph{prior probability} $\pr(\ud \theta)$ to the model parameters, with density also denoted $\pr(\theta)$.  Inference for the Bayesian model then consists of quantification of the \emph{posterior probability}
\begin{equation}\label{eq:bayes}
\pi(\theta)
\defeq
p(\theta| y)
=
\frac{L(\theta) \pr(\theta)}{p(y)},
\end{equation}
where the last equality, giving the posterior in terms of the likelihood and prior, is the practically useful formula of Bayes, and $p(y)$ is the \emph{model evidence}, defined by
$$
p(y) \defeq \int L(\theta) \pr(\theta) \ud \theta.
$$
\subsection{Challenges for inference}
In statistical models of practical interest, the likelihood $L(\theta)$ is often \emph{intractable}, meaning that it can not be evaluated pointwise.  However, in many settings which we consider, we will see that $L(\theta)$ can be estimated unbiasedly, meaning it is possible to generate a random variable $\hat{L}_\theta$ such that $\E[ \hat{L}_\theta ]=  L(\theta)$.  However, construction of a reliable unbiased estimator may be neither directly available, nor efficient.

The posterior $\pi(\theta)$ of the Bayesian model is in general intractable, and can not even be estimated unbiasedly.  This is often the case even if the likelihood is tractable, because of the normalisation by the model evidence in \eqref{eq:bayes}, which is usually computationally intensive to calculate.  In case of intractable likelihood in the Bayesian model, posterior inference is even more of a challenge, and one must usually rely on ergodic averages from Markov chain Monte Carlo (MCMC).  Such averages are generally asymptotically exact (i.e.~\emph{consistent}) as the number of iterations of the MCMC algorithm increases, in the sense of a law of large numbers.  However, MCMC can be computationally expensive to run.  It can take hours, days, weeks, or longer, in order to ensure a `reliable' MCMC estimate, where the level of reliability can be theoretically difficult to justify.


\subsection{Approximate families}
We will see that the use of approximations can help facilitate tractable, efficient and user-friendly inference.
Let $\mathscr{P}_\infty$ \label{eq:ideal-family} denote a set of (ideal) model probability distributions on $\Y$.
In many cases in practice, it may be desirable to work with a surrogate family of probability distributions $\mathscr{P}_0$.  Going further, it may be desirable to work with a family
\begin{equation}\label{eq:ell-approx-family}\nonumber
\mathscr{P}_\ell= \{p^{(\theta,\ell)} \}_{\theta\in\T},
\end{equation}
of data (probability) distributions, with $\ell \in [0,\infty]$ used to indicate families of increasingly `better' approximations.  For example, inference using $\mathscr{P}_\infty$ may be too difficult to achieve or too costly, in which case using an approximate family $\mathscr{P}_\ell$ may be possible instead.

It is conceivably possible to incorporate $\mathscr{P}_\ell$ in a Bayesian inference method, which may lessen the computational cost of the algorithm, while in the end performing inference for $\mathscr{P}_\infty$.  The aim of this thesis is to show general strategies in different settings where this is possible.

\subsection{Overview}\label{sec:intro-overview}
We now outline the remainder of this text\footnote{As for the intended audience, in order to keep the text of moderate size we must suppose some notions from analysis \citep[cf.][]{rudin}, probability \citep[cf.][]{georgii}, simulation \citep[cf.][]{liu-mc}, and statistics \citep[cf.][]{gelmanCSR}.  We try to strike a balance, to make the text of interest both to those knowledgeable and less knowledgeable in the subject matter of the thesis.}.  The text seeks to serve as an introduction and summary for the thesis papers listed on page \pageref{sec:list}. Methodological aspects are stressed for this introduction to the articles, as are  some of the supporting theoretical results. Most of the details are left to the articles.  For this introductory text, we do not give algorithms and results in full generality and for all cases. Rather we focus on a few important cases.  For example, we consider only a few specific Markov chains, rather than general Harris ergodic chains for the IS correction, and we focus on the use of unbiased estimators from particle filters\footnote{also known as \emph{sequential Monte Carlo}} in state space models, rather than from general importance sampling distributions in latent variable models.  Some more generality is provided in the original articles listed on page \pageref{sec:list}.   

We begin in Section \ref{sec:fk} with a specific problem of intractable likelihoods for statistical models, namely, that of the state space model, and review how interacting particle systems known as particle filters \cite{gordon-salmond-smith,stewart-mccarty} can lead to unbiased estimators of the $\infty$-likelihood (the likelihood corresponding to the family $\mathscr{P}_\infty$), as long as the dynamics of the state space model can be simulated.  We also detail an MCMC known as the particle marginal Metropolis-Hastings (PMMH) \cite{andrieuDH} (see also \cite{andrieu-roberts,beaumont,lin-liu-sloan,mollerPRB}), which allows for $\infty$-inference for the corresponding Bayesian model posterior, when unbiased $\infty$-likelihood estimators are available.

In Section \ref{sec:comparison}, as in \cite{viholaHF} we consider two different MCMCs, which are intended for acceleration of PMMH, and  which are based on use of an approximate family $\mathscr{P}_0$  and unbiased estimators for the $\infty$-likelihood.  These are the delayed acceptance (DA) MCMC \citep[cf.][]{lin-liu-sloan,banterle-grazian-lee-robert,christen-fox,cui-marzouk-willcox-scalable,golightlyHS,liu-mc} and the MCMC-IS \citep[cf.][]{doss,gilks-roberts,glynn-iglehart,hastings,parpasUWT,viholaHF}, both of which allow for unbiased estimators of the $0$-likelihood and $\infty$-likelihood, which can be useful when deterministic approximations are not available\footnote{The references \cite{golightlyHS} for DA and \cite{viholaHF} for MCMC-IS are most relevant in the unbiased estimator context for intractable likelihoods.}.  Based on an extension of the covariance ordering \cite{mira-geyer-ordering} to the IS setting, with differing reversible stationary probabilities and with unbiased estimators, we seek to compare the algorithms in terms of statistical efficiency, as in \citep[][]{franks-vihola}.

Section \ref{sec:diffusions} is concerned with a discretely and partially observed It\^{o} diffusion,  where unbiased $\infty$-likelihood estimates can not be directly obtained by the particle filter, because the dynamics of the diffusion can not be simulated.  Instead, approximate families $\mathscr{P}_\ell$ based on Euler approximation \citep[cf.][]{kloeden-platen} are used, along with multilevel \cite{heinrich,giles-or}, randomisation \cite{mcleish,rhee-glynn} and particle filter coupling \cite{jasraKLZ} techniques, leading to an unbiased estimator for the $\infty$-likelihood and to the Bayesian $\infty$-posterior by using an MCMC-IS with randomised weights, as in \cite{franksJLV}.   

Section \ref{sec:abc} is concerned with Bayesian models with intractable likelihoods, where an unbiased estimator of the likelihood is not readily available, but where it is at least possible to generate artificial observations from $p^{(\theta)}(\ud y')$.  The approach of approximate Bayesian computation \citep[cf.][]{sissonFB} is to use families $\mathscr{P}_{1/\epsilon}$ of approximations to $\mathscr{P}_\infty$, where $\mathscr{P}_{1/\epsilon}$ is indexed by $\epsilon>0$, the `tolerance,' which is difficult to choose.  We detail an approach based on an adaptive MCMC, as well as MCMC-IS \cite{wegmannLE}, with approximate confidence intervals for post-correction to finer tolerance, as in \cite{vihola-franks}.

We close with a brief discussion of ideas for future work in Section \ref{sec:discussion} and provide expanded individual summaries for the thesis papers in Section \ref{sec:summary}.

\section{Bayesian inference for state space models on path space}\label{sec:fk}
We introduce a well-known class of models based on latent variables on a state space $(\X,\mathcal{X})$ and conditionally independent observations on $(\Y,\mathcal{Y})$ which is sufficiently general and motivates a main application area of Articles \cite{viholaHF}, \cite{franks-vihola} and \cite{franksJLV} based on unbiased estimators and approximate families $\mathscr{P}_\ell$.

\subsection{Discretely-observed continuous-time path-dependent process}\label{sec:intro-fkm}
To motivate this general class of models, we consider an example of continuous-time latent process.
Suppose there is a process $(X_t')_{t\ge 0}$ of \emph{latent} or \emph{hidden states} $X_t'\in \X$, where $X_t'$ depends on $(X_s')_{0\le s <t}$ and the model parameter $\theta$.  Also, suppose $(Y_t')_{t\ge 0}$ is another process (of observations), where $Y_t'$ depends on $(X_s')_{0\le s\le t}$ and $\theta$.  
We make the realistic assumption\footnote{since the continuum can not easily be recorded by electronic or other physical means} that only finitely many observations $\{Y_{t_p}'\}_{p=0}^n$ are gathered at observation times $\{t_p\}_{p=0}^n$.

Let us set $Y_p \defeq Y_{t_p}'$ and $X_p\defeq X_{t_p}'$.
Let us define $X_{0:p}\defeq (X_0,\ldots, X_{p})$, and for
fixed parameter value $\theta$, consider the following dependency structure involving conditionally independent observations:
  \begin{equation*}
\begin{tikzcd}[ampersand replacement =\&]
  \cdots  \ar[r] \& X_{0:p-1} \ar[d] \ar{r} \& X_{0:p} \ar{d} \ar[r] \& X_{0:p+1} \ar[d] \ar[r] \& \cdots
  \\
  \& Y_{p-1} \& Y_p \& Y_{p+1} \&
  \end{tikzcd}
   \end{equation*}
  Here, the arrows denote a dependency relationship,  described in the following, where the initial state $X_0\sim \eta_0^{(\theta)}$ is drawn from an initial distribution $\eta_0^{(\theta)}$.  The dynamics between states (on path space) $X_{0:p-1}$ and $X_{0:p}$ is defined by a Markov probability kernel $\bar{M}_p^{(\theta,\infty)}$ (on path space),
where
$$
\bar{M}_p^{(\theta,\infty)}(x_{0:p-1}, \ud x_{0:p}')
\defeq
\charfun{x_{0:p-1}' = x_{0:p-1}}M_p^{(\theta,\infty)}(x_{0:p-1}, \ud x_p'),
$$
where $M_p^{(\theta,\infty)}$ is a Markov probability kernel from $\X^p$ to $\X$  induced by the dynamics of the path-dependent continuous-time process.  The observations $Y_p$ are obtained via $Y_p\sim g_p^{(\theta)}(\uarg| X_{0:p})$, where $g_p^{(\theta)}$ is the \emph{observational density}.

Let us set as shorthand $M_0^{(\theta,\infty)}(x_{0:-1},\ud x_0) \defeq \eta_0(\ud x_0)$ and
\begin{equation}\label{eq:fk-potential}\nonumber
  G_p^{(\theta)}(x_{0:p}) \defeq g_p^{(\theta)}(y_p| x_{0:p}).
  \end{equation}
The model described above in terms of the pair $(M_p^{(\theta,\infty)},G_p^{(\theta)})_{p=0}^n$ is known as a path-dependent \emph{state space model} (SSM)\footnote{As SSM is also known as a \emph{hidden Markov model} \citep[cf.][]{cappeMR}, especially in the engineering disciplines.}, or, more succinctly, as a \emph{Feynman-Kac model} \citep[cf.][]{del-moral}.

Simulation methods for the $\infty$-Feynman-Kac model are impossible if the dynamics $M_p^{(\theta,\ell)}$ can not be simulated exactly.  Besides some (important) exceptions, this is in general the case for continuous-time latent processes \citep[cf.][Sect.~1.3]{del-moral-mean}.  However, we will see when we consider It\^{o} diffusions in Section \ref{sec:diffusions}, that often one can obtain
Euler type approximations of the original process, with precision denoted `$\ell$', leading to approximate dynamics $M_p^{(\theta,\ell)}$ between observation times \citep[cf.][]{del-moral-mean,kloeden-platen}.  Using the same observational densities as for the exact model, we obtain a Feynman-Kac model $(M_p^{(\theta,\ell)}, G_p^{(\theta)})_{p=0}^n$ derived from the Euler type approximation of the dynamics.
  
\subsection{Model probabilities}\label{sec:intro-prob}
We now describe some of the probabilities associated to a Feynman-Kac model $(M_p^{(\theta,\ell)}, G_p^{(\theta)})_{p=0}^n$.

First, we define a bit of standard notation from analysis.  If $\mu$ is a probability measure and $s\ge 1$, we denote by $L^{s}(\mu)$ the Banach space of real-valued functions $\phi$, modulo equivalence in norm, with finite norm
\begin{equation}\label{eq:analysis-integral}
\mu(\abs{\phi}^s)^{\frac{1}{s}}<\infty,
\qquad\text{where}\qquad
\mu(\phi)\defeq \int \phi(x) \mu(\ud x).
\end{equation}

Consider now the conditional $\ell$-model probability on the latents, or \emph{$\ell$-smoother},
\begin{equation}\label{eq:latent-smoother}
p^{(\theta,\ell)}(\ud x_{0:n})
=
\frac{p^{(\theta,\ell)}_u(\ud x_{0:n})}{p^{(\theta,\ell)}_u(1)},
\end{equation}
where\footnote{In the notation $p_u^{(\theta,\ell)}(1)$, we view $1$ as the function $x_{0:n}\mapsto 1$, and the integral $p_u^{(\theta,\ell)}(\phi)=\int \phi(x_{0:n}) p_u^{(\theta,\ell)}(\ud x_{0:n})$ as in \eqref{eq:analysis-integral} for $\phi:\X^{n+1}\to\R$.}
  \begin{equation}\label{eq:ell-un-smoother}
  p^{(\theta,\ell)}_u(\ud x_{0:n})
  =  \bigg( \prod_{p=0}^n G^{(\theta)}_p(x_{0:p}) \bigg) \eta_0^{(\theta)}(\ud x_0) \prod_{p=1}^n M_p^{(\theta,\ell)}( x_{0:p-1}, \ud x_p). 
  \end{equation}
  Then $p^{(\theta,\ell)}$ represents the probability to observe the latent states given the observations according to the Feynman-Kac model  $(M_p^{(\theta,\ell)}, G_p^{(\theta)})_{p=0}^n$.
  In terms of a statistical model\footnote{really on $(\X\times\Y,\mathcal{X}\otimes\mathcal{Y})$, but we view $y_{0:n}\in \Y^{n+1}$ as fixed and therefore disregarded in the notation}
on $\X^{n+1}$, we have $\mathscr{P}_\ell=\{p^{(\theta,\ell)}\}_{\theta\in\T}$ with $p^{(\theta,\ell)}(\ud x_{0:n})$ defined in \eqref{eq:latent-smoother}, for the Feynman-Kac model.
  
The \emph{joint $\ell$-posterior probability} for the Bayesian model over model parameters and latent states is then 
  \begin{equation}\label{eq:ell-posterior}
  \pi^{(\ell)}(\ud \theta, \ud x_{0:n})
  \propto
  \pr(\ud \theta) p_u^{(\theta,\ell)}(\ud x_{0:n}).
  \end{equation}
  Writing the \emph{marginal} $\ell$-likelihood as $L^{(\ell)}(\theta)\defeq p_u^{(\theta,\ell)}(1)$ and considering the \emph{marginal} $\ell$-posterior on $\theta$, we obtain a more familiar formula to \eqref{eq:bayes} given in the beginning in Section \ref{sec:intro-bayesian}, namely,
  \begin{equation}\label{eq:marginal-posterior}\nonumber
  \pi_m^{(\ell)}(\ud \theta)
  =
  \int_{\X^{n+1}} \pi^{(\ell)}(\ud \theta, \ud x_{0:n})
  =
\frac{\pr(\ud \theta) L^{(\ell)}(\theta)}{\int \pr(\ud \theta) L^{(\ell)}(\theta)}.
  \end{equation}
The main topic of this thesis is incorporation of $\ell$-approximation within a $\infty$-inference method, to obtain efficient and user-friendly inference with respect to  $\pi^{(\infty)}$ and $\pi_m^{(\infty)}$.

  \subsection{Particle filter}\label{sec:intro-pf}
  Ignoring the $\theta$ and $\ell$ labels, we have seen that a Feynman-Kac model (with time horizon $n$) is defined through a pair $(M_p, G_p)_{p=0}^n$, where,
  \begin{enumerate}[(i)]
  \item $M_p(x_{0:p-1}, \ud x_p)$ is a Markov `transition' kernel for $p=1,\ldots, n$, and $M_0(x_{0:-1}, \ud x_0) \defeq \eta_0(\ud x_0)$ is a probability measure, and
    \item $G_p(x_{0:p})$ is a nonnegative `potential' function for $0\le p \le n$.
  \end{enumerate}


  The particle filter (PF) (Algorithm \ref{alg:pf}) was popularised in \citep[e.g.][]{stewart-mccarty,gordon-salmond-smith}, and allows for unbiased estimation \citep[cf.][]{del-moral,doucet-johansen} of
    \begin{equation}\label{eq:un-smoother}
  p_u(\ud x_{0:n})
  =
  \bigg( \prod_{p=0}^n G_p(x_{0:p})\bigg) \eta_0(\ud x_0) \prod_{p=1}^n M_p(x_{0:p-1}, \ud x_p),
  \end{equation}
 for (traditional) SSMs that are not path-dependent, that is,
\begin{align}\label{eq:traditional-ssm-one}
  M_p(X_{0:p-1}, \ud x_p') &= M_p(X_{p-1}, \ud x_p'),
  \\ \label{eq:traditional-ssm-two}
G_p(X_{0:p}) &= G_p(X_p).
\end{align}
However, straightforward generalisation also allows for unbiased estimators in the path-dependent setting of Feynman-Kac models, at least when the dynamics can be simulated \citep[cf.][]{del-moral}.  In addition, as is well-known, the general resampling scheme in PF (Algorithm \ref{alg:pf})  for ancestor random variables $\{A_p^{(i)}\}_{i=1}^N$ do lead to unbiased estimators, since the equality
$$
\E\Big[ \sum_{k=1}^N \charfun{A_p^{(k)} =i}\Big]
= N \frac{V_p^{(i)} }{V_p^*},
$$
is assumed satisfied for all $p\in\{0{:}n\}$ in PF (Algorithm \ref{alg:pf}) \citep[cf.][Prop.~$20$]{viholaHF}.
Such resampling schemes include  the popular multinomial, stratified, residual, and systematic resampling \citep[cf.][]{cappeMR,doucCM,doucet-johansen}.

The unbiased estimator, from the output $( V_n^{(i)},\mathbf{X}^{(i)})_{i=1}^n$ of PF (Algorithm \ref{alg:pf}) run for Feynman-Kac model $(M_p,G_p)_{p=0}^n$, is obtained by setting
  \begin{equation}\label{eq:p-hat}
    \hat{p}_u (\phi) \defeq\sum_{i=1}^N V_n^{(i)} \phi(\mathbf{X}^{(i)}),
    \end{equation}
 for $\phi\in L^1(p_u)$, which satisfies
\begin{equation}\label{eq:unbiased-estimation}
\E\big[ \hat{p}_u(\phi)\big]
  =p_u(\phi).
\end{equation}

An important point is that particle approximations $\hat{p}_u(\ud x_{0:n})$, for $p_u(\ud x_{0:n})$ through the PF for the model $(M_p, G_p)_{p=0}^n$, are not unique \citep[cf.][Sect.~2.4.2]{del-moral}.  One standard way to obtain a different particle approximation is merely changing the Feynman-Kac model to $(\tilde{M}_p,\tilde{G}_p)_{p=0}^n$, but in such a way that
  \begin{equation}\label{eq:smoothing-equal}
  \bigg( \prod_{p=0}^n \tilde{G}_p(x_{0:p})\bigg) \tilde{\eta}_0(\ud x_0) \prod_{p=1}^n \tilde{M}_p(x_{0:p-1}, \ud x_p)
  =
  \bigg( \prod_{p=0}^n G_p(x_{0:p})\bigg) \eta_0(\ud x_0) \prod_{p=1}^n M_p(x_{0:p-1}, \ud x_p).
  \end{equation}
holds, and running the PF (Algorithm \ref{alg:pf}) for the new Feynman-Kac model. 
From \eqref{eq:un-smoother} and \eqref{eq:unbiased-estimation}, it
follows that the unbiased estimator from the PF run for $(\tilde{M}_p, \tilde{G}_p)_{p=0}^n$ delivers the same unbiased estimation for $p_u(\ud x_{0:n})$ corresponding to the model $(M_p,G_p)_{p=0}^n$.
As an example for $(\tilde{M}_p, \tilde{G}_p)_{p=0}^n$, consider
$$
\tilde{G}_p(x_{0:p})
\defeq
\frac{ G_p(x_{0:p}) M_p(x_{0:p-1}, \ud x_p)}{\tilde{M}_p(x_{0:p-1}, \ud x_p)}(x_{0:p})
$$
in the sense of a Radon-Nikod\'{y}m derivative \citep[cf.][]{shiryaev}, which always exists if $M_p$ and $\tilde{M_p}$ admit densities and a support condition holds.

This non-uniqueness opens the door to consider more efficient PF implementations for a particular model and filtering/smoothing problem \citep[cf.][]{del-moral,doucet-johansen,guarnieroJL,pitt-shephard}.  The question of the optimal choice of $(M_p^\star,G_p^\star)_{p=0}^n$ for the smoothing problem (i.e.~unbiased estimation of $p_u(\ud x_{0:n})$) has been considered in \cite{guarnieroJL}.  As the optimal choice is usually not implementable, \cite{guarnieroJL} suggest an adaptive iterative algorithm, based on approximating families of mixtures of normals, in order to approximately find $M_p^\star$ and $G_p^\star$ (see also e.g.~\cite{hengBDD} for a related method).
Deterministic approximations, such as Laplace approximations \citep[cf.][]{rueMC}, can also be used as a substitute for the optimal transition $M_p^\star$ \cite{viholaHF}  (see also \cite{lindstenHV}).        
We emphasise that all the above mentioned approaches to the optimal choice problem achieve unbiased estimation of $p_u(\ud x_{0:n})$, as they use appropriately weighted potentials so that \eqref{eq:smoothing-equal} holds. 

  \begin{algorithm}
    \caption{Particle filter for Feynman-Kac model $(M_p,G_p)_{p=0}^n$, with $N\ge 1$ particles.}
    \label{alg:pf}
    \begin{flushleft}
      In the following, the particle index $i$ implicitly assumes all values in $\{1{:}N\}$.
    \end{flushleft}

    \begin{enumerate}[(1)]
\item For initialisation,
  \begin{enumerate}[(i)]
    \item
  Sample $X_0^{(i)} \sim \eta_0$.
  Set $\mathbf{X}^{(i)}\defeq X^{(i)}$.
  \newline
  Set $V_0^{(i)}\defeq \frac{1}{N} G_0(\mathbf{X}^{(i)})$ and set $V_0^* \defeq \sum_{j=1}^N V_0^{(j)}$.
\item
Sample random variables $\{A_0^{(k)}\}_{k=1}^N$ satisfying
        \newline$\E\big[ \sum_{k=1}^N \charfun{A_0^{(k)}=i}\big] = N V_0^{(i)} /V_0^*$.
    \end{enumerate}
\item For $p=1,\ldots n$,
  \begin{enumerate}[(i)]
    \item Sample
        $X_p^{(i)} \sim M_p(\mathbf{X}^{(A_{p-1}^{(i)})},\uarg)$.  Set $\mathbf{X}^{(i)}\defeq (\mathbf{X}^{(A_{p-1}^{(i)})}, X^{(i)})$.
\newline
         Set $V_p^{(i)}\defeq \big(V_{p-1}^*\big)\big( \frac{1}{N} G_p(\mathbf{X}^{(i)})\big)$  and set $V_p^* \defeq \sum_{j=1}^N V_p^{(j)}$.
\item
          Sample random variables $\{A_p^{(k)}\}_{k=1}^N$ satisfying
        \newline$\E\big[ \sum_{k=1}^N \charfun{A_p^{(k)}=i}\big] = N V_p^{(i)} /V_p^*.$
  \end{enumerate}
  \end{enumerate}
      \begin{flushleft}
        Output: $( V^{(i)},\mathbf{X}^{(i)})_{i=1}^N$, where $V^{(i)} \defeq V_n^{(i)}$.
        \end{flushleft}
  \end{algorithm}

Latent inference with respect to $p(\ud x_{0:n})$ is possible through the PF, at least when the dynamics $M_p$ can be simulated, by using a ratio estimator targeting $p_u(\phi)/p_u(1)$.  That is, if $\{\hat{p}_{u,k}\}_{k=1}^m$ with $m\ge 1$ are formed\footnote{Traditionally in particle filtering  \citep[cf.][]{doucet-johansen}, latent inference \eqref{eq:ratio-estimator} is done with $m=1$,  possibly with a  final resampling to form uniformly weighted particles, but final resampling leads to higher variance of the resulting estimator and is unnecessary here.} as in \eqref{eq:p-hat} from independent runs of PF (Algorithm \ref{alg:pf}) for $(M_p, G_p)_{p=0}^n$, then
  \begin{equation}\label{eq:ratio-estimator}
  \frac{ \sum_{k=1}^m \hat{p}_{u,k}(\phi)}{\sum_{k=1}^m \hat{p}_{u,k}(1)}
  \xrightarrow{m\to\infty}
  p^{(\theta)}(\phi),
  \qquad\text{almost surely}.
  \end{equation}
  
  We remark that the above estimator \eqref{eq:ratio-estimator}, as mentioned
for example in \citep[][Eq.~1]{chopin-central},
is an IS analogue of the `particle independent Metropolis-Hastings' (PIMH) \cite{andrieuDH} chain for latent smoothing.  The algorithm based on \eqref{eq:ratio-estimator} is completely parallelisable and does not depend on mixing of a chain, and is therefore relatively resilient in the number of particles $N$.
 Straightforward consistent estimators to construct confidence intervals are also available \citep[cf.][Prop.~23]{viholaHF}.
  
\subsection{Particle marginal Metropolis-Hastings}
The main task for which  we are interested is joint $\infty$-inference with respect to $\pi^{(\infty)}(\ud \theta, \ud x_{0:n})$.  So far, we only have shown how to perform $\infty$-inference for $p_u^{(\theta,\infty)}(\ud x_{0:n})$ and $p^{(\theta,\infty)}(\ud x_{0:n})$, with $\theta$ fixed.  Surprisingly \citep[cf.][]{andrieu-roberts,beaumont,lin-liu-sloan,mollerPRB}, joint inference is possible, using an MCMC known as the \emph{particle marginal Metropolis-Hastings} (PMMH) \cite{andrieuDH}.  
\begin{algorithm}
  \caption{Particle marginal Metropolis-Hastings, with $m\ge 1$ iterations.}
  \label{alg:pmmh}
  \begin{flushleft}
With $(\Theta_0, V_0^{(i)}, \mathbf{X}_0^{(i)})_{i=1}^N$ given, with $\sum_{i=1}^N V_0^{(i)}>0$, for $k=1,\ldots, m$, do:
    \end{flushleft}
  \begin{enumerate}[(i)]
  \item Sample $\Theta'\sim q(\uarg| \Theta_{k-1})$ from a transition kernel $q$ on $\T$.  
    \item
    Run PF (Algorithm \ref{alg:pf}) for $(M^{(\Theta', \infty)}, G^{(\Theta')})$, outputting $(V'^{(i)},\mathbf{X}'^{(i)})_{i=1}^N$.
    \item
    Accept, setting $(\Theta_k,V_k^{(i)}, \mathbf{X}_k^{(i)})_{i=1}^N \leftarrow (\Theta', {V'}^{(i)}, {\mathbf{X}'}^{(i)})_{i=1}^N$, with probability
    \begin{equation}\label{eq:acc-pmmh}
    \min\bigg\{
1, \frac{\pr(\Theta') \big( \sum_{i=1}^NV'^{(i)} \big) q(\Theta_{k-1}| \Theta')}{\pr(\Theta_{k-1})\big(\sum_{i=1}^N V_{k-1}^{(i)}\big) q(\Theta'| \Theta_{k-1})}
      \bigg\}.
      \end{equation}
      Otherwise, reject, setting $(\Theta_k,V_k^{(i)}, \mathbf{X}_k^{(i)})_{i=1}^N \leftarrow (\Theta_{k-1}, V_{k-1}^{(i)}, \mathbf{X}_{k-1}^{(i)})_{i=1}^N$.
\end{enumerate}
\end{algorithm}
Assuming $q(\theta'|\theta)>0$ for all $\theta,\,\theta'\in \T$ in the PMMH chain (Algorithm \ref{alg:pmmh}),
or a similarly mild condition ensuring Harris ergodicity of the chain \citep[cf.][]{meyn-tweedie},
the estimator formed from PMMH is strongly consistent:  for $f\in L^1(\pi^{(\infty)})$,
\begin{equation}\label{eq:pmmh-estimator}
E_m^{PM}(f)
\defeq \frac{1}{m} \sum_{k=1}^m  \sum_{i=1}^N \frac{V_k^{(i)} f(\Theta_k, \mathbf{X}_k^{(i)})}{\sum_{j=1}^N V_k^{(j)}} 
\xrightarrow{m\to\infty}
\pi^{(\infty)}(f),
\qquad\text{a.s.}\footnote{almost surely}
\end{equation}

We remark about `Metropolis-Hastings type' MCMC.
The PMMH \cite{andrieuDH} is used in state space models using PFs, \emph{pseudo-marginal} MCMC \cite{andrieu-roberts,beaumont,lin-liu-sloan,mollerPRB} is the general term for the chain used in latent variable models with unbiased estimators,  and Metropolis-Hastings MCMC \citep[][]{metropolisRRT,hastings} is used in Bayesian models with tractable likelihoods.  In fact, it is possible to view these `Metropolis-Hastings type' MCMCs each as a substantiation of the other: one direction follows by viewing the pseudo-marginal MCMC and PMMH as full-dimensional Metropolis-Hastings kernels on an extended state space, while the other direction follows by trivialisation \citep[cf.][]{andrieu-roberts}.
\section{Accelerations based on an approximation}\label{sec:comparison}
The Metropolis-Hastings MCMC has served as the backbone of the MCMC revolution for half of the last century \cite{diaconis-revolution}, while pseudo-marginal MCMC and the PMMH have been quite popular and extensively used in the current century (see \citep[Sect.~$1.2$]{franks-vihola} for a review).  Because of the importance of these MCMCs, there has been considerable interest in their possible acceleration.  We focus on acceleration of the PMMH in the following.

Usually by far the most computationally intensive part of the PMMH is running the PF (Algorithm \ref{alg:pf}), for $(M_p^{(\theta,\infty)}, G_p^{(\theta)})_{p=0}^n$ with output $(V^{(i)},\mathbf{X}^{(i)})_{i=1}^N$, to obtain the unbiased estimator
\begin{equation}\label{eq:likelihood-estimator}\nonumber
\hat{L}^{(\infty)}(\theta) \defeq \hat{p}_u^{(\theta,\infty)}(1)
= \sum_{i=1}^N V^{(i)}
\end{equation}
of the likelihood $L^{(\infty)}(\theta)$.
The idea of acceleration based on approximation  is to substitute a computationally cheaper (non-negative unbiased estimator $\hat{L}^{(0)}(\theta)$ of an) approximation $L^{(0)}(\theta)$ for the $\infty$-likelihood, instead of using $\hat{L}^{(\infty)}(\theta)$.  One would also like to maintain (strong) consistency of the resulting estimator for the $\infty$-posterior.

\subsection{Delayed acceptance and importance sampling}
One such popular acceleration algorithm is delayed acceptance (DA) (Algorithm \ref{alg:da}) \citep[cf.][]{lin-liu-sloan,banterle-grazian-lee-robert,christen-fox,cui-marzouk-willcox-scalable,golightlyHS,liu-mc}, with $\epsilon \ge 0$.  
\begin{algorithm}
  \caption{Delayed acceptance, with $m\ge 1$ iterations, and $\epsilon\ge 0$}
  \label{alg:da}
  \begin{flushleft}
    Given $(\Theta_0, V_0^{(i)}, \mathbf{X}_0^{(i)}, \hat{L}^{(0)}(\Theta_0))_{i=1}^N$, with $\sum_{i=1}^N V_0^{(i)}>0$ and $\hat{L}^{(0)}(\Theta_0)>0$.
    \newline
    For $k=1,\ldots, m$, do:
    \end{flushleft}
  \begin{enumerate}[(i)]
  \item\label{alg:da-one} Sample $\Theta'\sim q(\uarg| \Theta_{k-1})$ from a transition kernel $q$ on $\T$.  
    \newline
    Obtain unbiased estimate $\hat{L}^{(0)}(\Theta')$ of  $L^{(0)}(\Theta')$.
    \newline
    Proceed to step \eqref{alg:da-two} with probability
    \begin{equation}\label{eq:acc-da-one}
    \min\bigg\{
1, \frac{\pr(\Theta')\big(\hat{L}^{(0)}(\Theta')+\epsilon\big) q(\Theta_{k-1}| \Theta')}{\pr(\Theta_{k-1})\big(\hat{L}^{(0)}(\Theta_{k-1})+\epsilon\big) q(\Theta'| \Theta_{k-1})}
\bigg\}.
\end{equation}
    Otherwise, reject, setting
    \newline
    $(\Theta_k,V_k^{(i)},\mathbf{X}_k^{(i)},\hat{L}^{(0)}(\Theta_k) )_{i=1}^N\leftarrow (\Theta_{k-1}, V_{k-1}^{(i)}, \mathbf{X}_{k-1}^{(i)}, \hat{L}^{(0)}(\Theta_{k-1}))_{i=1}^N$.
\item\label{alg:da-two}
    Run PF (Algorithm \ref{alg:pf}) for $(M^{(\Theta', \infty)}, G^{(\Theta')})$, outputting $(V'^{(i)},\mathbf{X}'^{(i)})_{i=1}^N$.
    Accept, setting $(\Theta_k, V_k^{(i)}, \mathbf{X}_k^{(i)}, \hat{L}^{(0)}(\Theta_k) )_{i=1}^N \leftarrow (\Theta', {V'}^{(i)},{\mathbf{X}'}^{(i)}, \hat{L}^{(0)}(\Theta'))_{i=1}^N$, with probability
    \begin{equation}\label{eq:acc-da-two}
    \min\bigg\{
1, \frac{\big(\sum_{i=1}^N V'^{(i)}\big)/\big(\hat{L}^{(0)}(\Theta')+\epsilon\big)}{\big( \sum_{i=1}^N V_{k-1}^{(i)}\big)/\big(\hat{L}^{(0)}(\Theta_{k-1})+\epsilon\big)}
      \bigg\}.
      \end{equation}
    Otherwise, reject, setting
    \newline
    $(\Theta_k, V_k^{(i)}, \mathbf{X}_k^{(i)}, \hat{L}^{(0)}(\Theta_k))_{i=1}^N\leftarrow (\Theta_{k-1}, V_{k-1}^{(i)}, \mathbf{X}_{k-1}^{(i)}, \hat{L}^{(0)}(\Theta_{k-1}))_{i=1}^N$.
\end{enumerate}
\end{algorithm}
We require that almost surely the support condition
\begin{equation}\label{eq:support-condition}
\hat{L}^{(\infty)}(\theta) >0 \implies \big(\hat{L}^{(0)}(\theta)+\epsilon\big)>0
\end{equation}
holds, so that the resulting weight $\hat{L}^{(\infty)}(\theta)/\big(\hat{L}^{(0)}(\theta)+\epsilon\big)$ in Algorithm \ref{alg:da}\eqref{alg:da-two} is guaranteed well-defined.
This can be simply achieved always by choosing a regularisation constant\footnote{This will be done in Algorithm \ref{alg:is-diffusion} given later, and is linked to `defensive importance sampling' \cite{hesterberg}.} $\epsilon>0$, leading to asymptotically exact $\infty$-inference.
We note that step \eqref{alg:da-one} in DA (Algorithm \ref{alg:da}) effectively acts as a screening stage: only `good' proposals proceed to step \eqref{alg:da-two}, where the expensive $\infty$-model PF must be run. The resulting DA estimator for the $\infty$-posterior is the same as that of PMMH, given in \eqref{eq:pmmh-estimator}.

As an alternative to PMMH/DA, we consider MCMC-IS (Algorithm \ref{alg:is})
\citep[cf.][]{doss,gilks-roberts,glynn-iglehart,hastings,parpasUWT,viholaHF}.
Here, for $f:\T\times \X^{n+1}\rightarrow \R$ we have set $f^{(\theta)}(x_{0:n}) = f(\theta,x_{0:n})$.  
\begin{algorithm}
  \caption{MCMC-IS.  Importance sampling correction of PMMH, with $m\ge 1$ iterations, and $\epsilon\ge 0$.}
  \label{alg:is}
  \begin{enumerate}[(P1)]
  \item\label{alg:is-one}
 Given $(\Theta_0, \hat{L}^{(0)}(\Theta_0))$, with $\hat{L}^{(0)}(\Theta_0)>0$, for $k=1,\ldots, m$, do:
    \begin{enumerate}[(i)]
\item
  Sample $\Theta'\sim q(\uarg| \Theta_{k-1})$ from a transition kernel $q$.  
\item
    Obtain unbiased estimate $\hat{L}^{(0)}(\Theta')$ of $L^{(0)}(\Theta')$.
\item       Accept, setting $(\Theta_k, \hat{L}^{(0)}(\Theta_k)) \leftarrow (\Theta',\hat{L}^{(0)}(\Theta'))$, with probability \eqref{eq:acc-da-one}. 
Otherwise, reject, setting $(\Theta_k, \hat{L}^{(0)}(\Theta_k)) \leftarrow (\Theta_{k-1}, \hat{L}^{(0)}(\Theta_{k-1}))$.
\end{enumerate}
  \item\label{alg:is-two}
    For all $k\in \{1{:}m\}$,
    \begin{enumerate}[(i)]
 \item Run PF (Algorithm \ref{alg:pf}) for $(M^{(\Theta_k, \infty)}, G^{(\Theta_k)})$, outputting $(V_k^{(i)},\mathbf{X}_k^{(i)})_{i=1}^N$.
\item   Set
  $
  \xi_k(\phi) \defeq \frac{\sum_{i=1}^N V_k^{(i)} \phi(\mathbf{X}_k^{(i)})}{\hat{L}^{(0)}(\Theta_k)+\epsilon},
  $
  for $\phi:\X^{n+1}\rightarrow \R$. 
     Form the estimator,
  \begin{equation}\label{eq:is-estimator}
  E_m^{IS}(f)\defeq
  \frac{\sum_{k=1}^m \xi_k(f^{(\Theta_k)})}{\sum_{k=1}^m \xi_k(1)}.
  \end{equation}
\end{enumerate}
    \end{enumerate}
\end{algorithm}
Assuming the Phase \ref{alg:is-one} chain is Harris ergodic (e.g.~$q(\theta,\theta')>0$ for all $\theta,\theta'\in\T$) and the support condition \eqref{eq:support-condition} holds, like the PMMH/DA estimator, the MCMC-IS estimator is strongly consistent \citep[][Thm.~3]{viholaHF}: for $f\in L^1(\pi^{(\infty)})$,
$$
E_m^{IS}(f)
\xrightarrow{m\rightarrow\infty} \pi^{(\infty)}(f),
\qquad\text{almost surely}.
$$
Phase \ref{alg:is-one} of MCMC-IS (Algorithm \ref{alg:is-one}) implements a PMMH (Algorithm \ref{alg:pmmh}) targeting marginally 
$$
\pi_m^{(0)}(\theta) \propto \pr(\theta) L^{(0)}(\theta).
$$
Phase \ref{alg:is-two} consists of independent calls of PF (Algorithm \ref{alg:pf}), and is therefore completely parallelisable, unlike DA (Algorithm \ref{alg:da}).  This allows for the possibility of substantial additional speedup on a parallel computer \citep[cf.][]{leeYGDH}.

We remark about an acceleration technique known as `early rejection' \cite{solonenOLHTJ} for Metropolis-Hastings, that can sometimes be employed if the likelihood takes a special form, described below.\footnote{A similar idea of early cancellation as `early rejection' has been used previously in the exact simulation literature, under the name of `retrospective simulation' \citep[][]{beskosPR}.} 
The acceleration technique  also applies to DA step \eqref{alg:da-one} and MCMC-IS Phase \ref{alg:is-one}, if $\hat{L}^{(0)}(\theta) =L^{(0)}(\theta)$ almost surely and $\epsilon=0$.  The form required in \cite{solonenOLHTJ} is that the $0$-likelihood $L^{(0)}(\theta)$  can be written, for example, as
$$
L^{(0)}(\theta) 
\propto
\prod_{j=0}^n \exp\big( -l_j^{(\theta,0)}(y_j)\big)
$$
with $l_j^{(\theta,0)}(y_j)\ge 0$.  In this case, because the likelihood only gets smaller with more components of the product computed, the calculation of the components can be ended and the proposal rejected early in acceptance probability \eqref{eq:acc-da-one} for DA and MCMC-IS, as soon as the partially computed acceptance probability in \eqref{eq:acc-da-one} becomes smaller than the
uniformly generated random variable \citep[cf.][Sect. 4]{solonenOLHTJ}.  The `early rejection' trick requires a special form for the likelihood, however, and therefore is not always applicable.

\subsection{The question of relative efficiency}
The delayed acceptance and importance sampling correction are two acceleration alternatives to the standard PMMH, both of which use the same approximation and algorithmic ingredients.  The question of choice of alternative methods has been remarked before \cite{cui-marzouk-willcox-scalable} in the simpler setting of Metropolis-Hastings, without unbiased estimators.
Article \cite{viholaHF} introduces the IS correction in the general case of unbiased estimators in both Phase \ref{alg:is-one} and Phase \ref{alg:is-two}, and seeks to compare MCMC-IS with DA in the general setting.

A numerical comparison of the methods is done in \cite{viholaHF}, where the MCMC-IS approach was found to work slightly better than DA in experiments in SSMs, even without parallelisation.  As an example of a computationally intensive experiment, a stochastic volatility model was considered with observation consisting of real data from daily financial index returns spanning two decades.
Laplace approximations were used to approximate the $0$-likelihood, and were used as well in the IS correction, namely, for the approximation to the optimal choice\footnote{as discussed in Section \ref{sec:intro-pf}} of Feynman-Kac model for the smoothing problem for $p_u^{(\theta,\infty)}(\ud x_{0:n})$.
With all methods making intelligent use of the Laplace approximations, MCMC-IS performed significantly better than PMMH or DA in the experiment.

In additional to the experiments, many additional potential enhancements were suggested in \cite{viholaHF} which would improve the computational efficiency of MCMC-IS in practice, relative to DA acceleration of PMMH, even further.  For example, the Phase \ref{alg:is-two} IS weights do not need to be calculated during the burn-in phase\footnote{Additionally, the debiasing tricks \citep[cf.][]{glynn-rhee} may be effectively and efficiently used.}  and for thinned out samples of the chain\footnote{Thinning  \citep[cf.][]{owen-thinning} denotes the procedure, in which only every $k^{th}$ sample of the Markov chain is kept, with say $k=10$, in order to decrease the auto-correlation of samples.}, nor for repeated samples of the chain if the jump chain\footnote{the chain formed formed from the original chain, consisting only of the accepted states of the original chain \citep[cf.][]{douc-robert,viholaHF}} is used.  As well, as previously mentioned, Phase \ref{alg:is-two} admits a straightforward parallelisation for calculation of the more expensive IS weights, which significantly increases the scalability and efficiency of MCMC-IS.

\subsection{Peskun and covariance orderings of asymptotic variances}\label{sec:comparison-asvar}
An estimator $E_m(f)$ is said to satisfy a central limit theorem (CLT), if
\begin{equation}\label{eq:clt}\nonumber
\sqrt{m}\big[
E_m(f) - \pi^{(\infty)}(f)
\big]
\xrightarrow{m\to\infty}
\mathrm{N}\big(0, \sigma^2(f)\big),
\qquad
\text{in distribution}.
\end{equation}
In this case, we call $\sigma^2(f)$ the \emph{asymptotic variance} of the estimator. 

Without taking into account computational factors previously mentioned (which generally support the use of MCMC-IS; see also Section \ref{sec:diffusions-efficiency}), and considering just the statistical efficiency of the estimators in terms of the asymptotic variance, it was found in \cite{franks-vihola} through artificially constructed toy examples that either MCMC-IS or PMMH/DA may do arbitrary better than the other.  Moreover, the examples seemed to indicate that MCMC-IS might do better in cases of practical interest, with multi-modal targets, a phenomenon remarked previously about MCMC-IS and Metropolis-Hastings \citep[e.g.][]{gilks-roberts}.
Proving that the IS acceleration is usually `better' than DA is of course a separate matter, which can not be done based on experiments or examples alone.

We first introduce some notation and terminology.  A Markov kernel $K$ on $(\X, \mathcal{X})$ is said to be \emph{reversible} with respect to a probability $\mu$, if for all $A,B\in \mathcal{X}$, 
$$
\int \mu(\ud x) K(x, \ud y) \charfun{x\in A, \, y\in B} = \int \mu(\ud y) K(y, \ud x) \charfun{x\in A,\, y\in B}.
$$
We also define the \emph{Dirichlet form}
\begin{equation}\label{eq:dirichlet-form}\nonumber
\cE_K(g) \defeq \inner{g}{(1-K)g}_\mu
\end{equation}
for $g\in L^2(\mu)$, 
where $\inner{g_1}{g_2}_\mu \defeq \int g_1(x) g_2(x) \mu(\ud x)$,
$Kg(x)\defeq \int K(x,\ud x') g(x')$ and $(1 g)(x) = g(x)$.

The famous Peskun ordering \cite{peskun,tierney-note} says that if
\begin{equation}\label{eq:off-diagonal}
K(x,A\backslash\{x\}) \ge L(x,A\backslash\{x\})
\qquad \text{$\mu$-almost every } x\in \X,\,\forall A\in \mathcal{X},
\end{equation}
where $K$ and $L$ are two Markov kernels, both reversible with respect to a probability $\mu$, then
\begin{equation}\label{eq:peskun}
\sigma_K^2(f)\le \sigma_L^2(f)
\qquad\forall f\in L^2(\mu),
\end{equation}
where $\sigma_K^2(f)$ and $\sigma_L^2(f)$ denote the asymptotic variances of the $K$ and $L$ chains, respectively.

Consider next a popular Peskun `type' comparison result for asymptotic variances of reversible chains, known as the covariance ordering\footnote{Though not mentioned by name, it was shown already in \citep[][Proof of Lem.~3]{tierney-note} that the Peskun ordering is equivalent with the `covariance' ordering.}  \cite{mira-geyer-ordering}: if $K$ and $L$ are two Markov kernels, both reversible with respect to a probability $\mu$, and if
\begin{equation}\label{eq:covariance-dirichlet}
\cE_K(g)\ge \cE_L(g),
\qquad\forall g\in L^2(\mu),
\end{equation}
then
\begin{equation}\label{eq:covariance}
  \sigma_K^2(f) \le \sigma_L^2(f)
  \qquad\forall f\in L^2(\mu).
\end{equation}
Compared to the Peskun ordering, the covariance ordering can be more useful in practice, as the criterion can distinguish better between chains on general state spaces.  For example, some chains vanish along the diagonal, in which case \eqref{eq:off-diagonal} may be useless, but \eqref{eq:covariance-dirichlet} may still be able to distinguish between these chains \citep[cf.][]{mira-geyer-ordering,mira-leisen-covariance}.

As a simple application of the covariance ordering, let us consider the case of PMMH and DA, which are both reversible with respect to the same invariant measure (see \citep[][]{banterle-grazian-lee-robert} or Section \ref{sec:comparison-guarantees}).
Using the identity
$$
\cE_L(g)
=
\frac{1}{2}
\int \mu(\ud x) L(x, \ud y) \big(g(x) - g(y)\big)^2,
$$
which holds for any $\mu$-reversible kernel $L$, and
that the product of the acceptance probabilities \eqref{eq:acc-da-one} and \eqref{eq:acc-da-two} in DA (Algorithm \ref{alg:da}) is less than or equal to the acceptance probability \eqref{eq:acc-pmmh} in PMMH (Algorithm \ref{alg:pmmh}), it can be shown \citep[cf.][]{banterle-grazian-lee-robert} that the covariance ordering implies
$$
\sigma_{PM}^{2}(f) \le \sigma_{DA}^{2}(f).
$$

\subsection{Peskun type ordering for importance sampling correction}\label{sec:comparison-peskun-type}
Article \cite{franks-vihola} is concerned with extending the covariance ordering to chains $K$ and $L$ reversible with respect to probabilities $\Pi^{(0)}$ and $\Pi^{(\infty)}$, where $\Pi^{(0)}$ and $\Pi^{(\infty)}$ may be different.  

Suppose then that $K$ and $L$ are Harris ergodic chains on a space $(\X,\mathcal{X})$, where $K$ is $\Pi^{(0)}$-reversible and $L$ is $\Pi^{(\infty)}$-reversible.  Suppose further that the Radon-Nikod\'{y}m derivative\footnote{This is the function $w$ satisfying $\Pi^{(0)}( w g) = \Pi^{(\infty)}(g)$ for all $g\in L^1(\Pi^{(\infty)})$.}
\begin{equation}\label{eq:importance-sampling-weight}\nonumber
w(x) \defeq \frac{\ud \Pi^{(\infty)}}{\ud \Pi^{(0)}}(x)
\end{equation}
exists.  Let $\lowc,\,\highc \ge 0$ be constants such that

\begin{align*}
\lowc\, \cE_K(g) \le &\cE_L(g) \le \highc\, \cE_K(g)\\
\lowc \,\le &w(x) \le \,\highc,
\end{align*}
for all $x\in \X$ and $g\in L^2(\Pi^{(0)})$.  Then \citep[Thm.~2]{franks-vihola}, for all $f\in L^2(\Pi^{(\infty)})$ with $\bar{f}\defeq f - \Pi^{(\infty)}(f)$, we have
\begin{align}\label{eq:covariance-is-one}
  \sigma_K^2(f) + \var_{\Pi^{(0)}}(w \bar{f}) &\le \highc\Big( \sigma_L^2(f) + \var_{\Pi^{(\infty)}}( f)\Big), \\ \label{eq:covariance-is-two}
  \sigma_K^2(f) + \var_{\Pi^{(0)}}(w \bar{f}) &\ge \lowc\Big( \sigma_L^2(f) + \var_{\Pi^{(\infty)}}( f)\Big). 
\end{align}
If $\Pi^{(0)} = \Pi^{(\infty)}$, then it is direct to see that \eqref{eq:covariance-is-one} simplifies to the covariance ordering \eqref{eq:covariance} given earlier.  Versions of the orderings (\ref{eq:covariance-is-one}-\ref{eq:covariance-is-two}) also hold for when the marginal weight is bounded in a latent variable setting \citep[Thm.~5]{franks-vihola}, and for self-normalised estimators using jump chain representation and unbiased estimators \citep[Thm.~12]{franks-vihola} to compare with pseudo-marginal type MCMC.  We discuss a particular implication of these orderings in the next section, namely MCMC-IS (algorithm \ref{alg:is}) compared to PMMH (Algorithm \ref{alg:pmmh}) and DA (Algorithm \ref{alg:da}).   

\subsection{Comparison results}\label{sec:comparison-guarantees}
We are now ready to compare MCMC-IS (Algorithm \ref{alg:is}) with PMMH (Algorithm \ref{alg:pmmh}) and DA (Algorithm \ref{alg:da}) in terms of the asymptotic variance.
For simplicity, we assume deterministic approximation for the $0$-likelihood, that is, $\hat{L}^{(0)}(\theta) = L^{(0)}(\theta)$ almost surely.\footnote{For the general case for $\hat{L}^{(0)}(\theta)$, see \citep[Thm.~14]{franks-vihola}.}  
We note that the MCMC-IS chain is $\Pi^{(0)}$-reversible, while the PMMH and DA chains are both $\Pi^{(\infty)}$-reversible, with probabilities defined in the following.

Article \cite{franks-vihola} shows how a comparison can be made when the (marginal) weight between the approximate and exact model posteriors $w$ (or $\wmarg$) is bounded (the weights $w$ and $\wmarg$ are defined below).  This follows from  the extension of the covariance ordering to the IS context with unbiased estimators, mentioned earlier.

We first need to define some notation.  Let
$
Q_\theta^{(\infty)}(\ud \bx^{(1:N)}, \ud v^{(1:N)})
$
denote the law of the output $(\bX^{(1:N)}, V^{(1:N)})$ of the PF (Algorithm \ref{alg:pf}) for the model $(M_p^{(\theta,\infty)}, G_p^{(\theta)})_{p=0}^n$.
The full invariant probability of the PMMH (Algorithm \ref{alg:pmmh}) is then given by 
$$
\Pi^{(\infty)}(\ud \theta, \ud v^{(1:N)},\ud \bx^{(1:N)})
=
\frac{1}{\const{\infty}}
\pr(\ud \theta) Q_\theta^{(\infty)}(\ud \bx^{(1:N)}, \ud v^{(1:N)}) \sum_{i=1}^N v^{(i)},
$$
where $\const{\infty}$ is a normalising constant.
The full invariant probability of the IS corrected chain (Algorithm \ref{alg:is}) is 
$$
\Pi^{(0)}(\ud \theta,  \ud v^{(1:N)},\ud \bx^{(1:N)})
=
\frac{1}{\const{0}}
\pr(\ud \theta)\big( L^{(0)}(\theta)+\epsilon\big) Q_\theta^{(\infty)}(\ud \bx^{(1:N)}, \ud v^{(1:N)}),
$$
where $\const{0}$ is a normalising constant. 
We set for a function $f:\T\times\X^{n+1}\to\R$, 
$$
\hat{\zeta}(f)
\defeq
\frac{\zeta(f)}{\zeta(1)},
\qquad\text{where}\qquad
\zeta(f)\defeq \sum_{i=1}^N V^{(i)} f(\Theta, \mathbf{X}^{(i)}).
$$
Assuming 
\begin{equation}\label{eq:support-condition-simple}
\hat{L}^{(\infty)}(\theta)>0 \implies \big( L^{(0)}(\theta)+\epsilon\big)>0
\end{equation}
almost surely, for some $\epsilon \ge 0$,
the  weights
$$
w(\theta,  v^{(1:N)},\mathbf{x}^{(1:N)} ) \defeq
\frac{\const{0} }{\const{\infty} } \frac{1}{L^{(0)}(\theta)+\epsilon} \sum_{i=1}^N v^{(i)},  
\qquad\text{and}\qquad
\wmarg(\theta) \defeq
\frac{c_0}{c_\infty}\frac{ L^{(\infty)}(\theta) }{L^{(0)}(\theta)+\epsilon},
$$
correspond to the Radon-Nikod\'{y}m derivatives between the approximate and exact model full and marginal posteriors. 

Let us now describe a CLT for MCMC-IS.
As before, for a function $f\in L^2(\pi^{(\infty)})$, we set $\bar{f} \defeq f- \pi^{(\infty)}(f)$.  By \citep[Thm.~7(i)]{viholaHF}, for $f\in L^2(\pi^{(\infty)})$ the MCMC-IS estimator \eqref{eq:is-estimator} satisfies a CLT, with
a formula for the MCMC-IS asymptotic variance given by
\begin{equation}\label{eq:is-asvar}
\sigma^2_{IS}(f)
=
\sigma^2_{IS,1}(f) + \sigma^2_{IS,2}(f),
\end{equation}
assuming $\sigma_{IS}^2(f)<\infty$,
 support condition \eqref{eq:support-condition} holds, and the marginal chain $(\Theta_k)_{k\ge 1}$ of MCMC-IS (Algorithm \ref{alg:is}) is Harris ergodic\footnote{E.g.~$q(\theta,\theta')>0$ for all $\theta,\theta'\in\T$.}  and aperiodic\footnote{See for example \cite{meyn-tweedie} for this and other definitions.}.  Here, $\sigma_{IS,1}^2(f)$ 
is the asymptotic variance of the marginal chain $(\Theta)_{k\ge 1}$, that is,
$$
\frac{1}{\sqrt{m}} \sum_{k=1}^m \E\big[ w(\theta,  V^{(1:N)},\mathbf{X}^{(1:N)})  \hat{\zeta}(\bar{f})| \Theta_k=\theta\big]
\xrightarrow{m\to \infty}
\mathrm{N}\big(0,\sigma_{IS,1}^2(f)\big),
$$
and
$$
\sigma_{IS,2}^2(f)\defeq \pi_m^{(0)}\big( v_{w \hat{\zeta}(\bar{f})}\big),
$$
with
$$
v_g(\theta)\defeq \var\big( g(\theta, V^{(1:N)},\mathbf{X}^{(1:N)})|\Theta_k = \theta \big). 
$$
Note the decomposition of the MCMC-IS asymptotic variance \eqref{eq:is-asvar} into marginal MCMC and IS correction components, which may be helpful in questions of tuning and allocation of computational resources.  A similar decomposition is not expected to hold for the DA asymptotic variance.

We now state the comparison results between MCMC-IS and PMMH/DA.  For functions $f\in L^2(\pi^{(\infty)})$, such that the CLT and conditions given above for MCMC-IS hold, and assuming the PMMH and DA chains are Harris ergodic, we have the following comparison result \citep[Thm.~14]{franks-vihola}, with $\sigma_L^2$ equal to $\sigma_{PM}^2$ or $\sigma_{DA}^2$:
\begin{equation}\label{eq:is-da-marginal}
\sigma_{IS}^2(f) \le
(\sup \wmarg) \Big( \sigma_{L}^2(f) + \var_{\Pi^{(\infty)}}\big( \hat{\zeta}(f)\big) \Big) + 3 \var_{\Pi^{(0)}}\big( w \hat{\zeta}(\bar{f})\big).
\end{equation}
Note that $\sup \wmarg \le \sup w$.  We have, moreover, if also $\sup w<\infty$, 
\begin{align}
  \label{eq:is-da-full}
  \sigma_{IS}^2(f) + \var_{\Pi^{(0)}}\big(w \hat{\zeta}(\bar{f}) \big)
  &\le
  (\sup w)\Big( \sigma_{L}^2(f) + \var_{\Pi^{(\infty)}}\big( \hat{\zeta}(f)\big)\Big)
    \\ \label{eq:is-da-reverse}
   \sigma_{IS}^2(f) + \var_{\Pi^{(0)}}\big(w \hat{\zeta}(\bar{f}) \big)
  &\ge
  (\inf w)\Big( \sigma_{L}^2(f) + \var_{\Pi^{(\infty)}}\big( \hat{\zeta}(f)\big)\Big)
\end{align}

The results show that the asymptotic variance of MCMC-IS and PMMH/DA can be related up to additive and multiplicative constants, which can be informative by \eqref{eq:is-da-marginal} in practical cases where the marginal weight $\wmarg$ is bounded, where $\wmarg$ relates the ratio of likelihoods.  We note that \eqref{eq:is-da-reverse} is usually not helpful since a positive lower bound on $w$  is usually not possible, while \eqref{eq:is-da-marginal} and \eqref{eq:is-da-full} do not require a positive lower bound, and are therefore more generally applicable, providing theoretical guarantees for MCMC-IS in terms of PMMH/DA.
Another  nice facet of (\ref{eq:is-da-marginal}-\ref{eq:is-da-reverse}) is that the function $f\in L^2(\Pi^{(\infty)})$ is allowed to be a function on $\T\times\X^{n+1}$, not only on $\T$.

Also shown in \cite{franks-vihola} is the not too surprising fact that geometric ergodicity of the MCMC-IS augmented chain is inherited by its marginal chain.  This relates the fact that the convergence and mixing of the MCMC-IS chain is not affected by the noise in the Phase \ref{alg:is-two} unbiased estimators, unlike PMMH and DA, which are very dependent on the noise, and are not geometrically ergodic if the unbiased estimator is unbounded \citep[cf.][]{andrieu-roberts,franks-vihola}.  Of course, the asymptotic variance \eqref{eq:is-asvar} of the MCMC-IS estimator \eqref{eq:is-estimator} depends on the noise, but it seems it is not as harmful in the output estimator compared to in the acceptance ratios \eqref{eq:acc-pmmh} and \eqref{eq:acc-da-two} of PMMH and DA, respectively.  Besides convergence and mixing, geometric ergodicity is also likely helpful for example in estimation of the asymptotic variance \citep[cf.][]{flegal-jones}, as well as in verifying convergence of adaptive MCMC schemes \citep[cf.][]{andrieu-thoms}, at least based on the existing theory.  

There is room for further theoretical development.  For example,
quantification of the error of MCMC-IS and of the asymptotic variance, could be investigated along the lines of \cite{flegal-jones,royTF}.  Also, in terms of non-asymptotic error bounds, results for MCMC \citep[e.g.][]{latuszynskiMN,mathe-novak,rudolf-explicit} could likely be extended to MCMC-IS.

\section{Bayesian inference for state space models with diffusion dynamics}\label{sec:diffusions}

The PF (Algorithm \ref{alg:pf}) for the Feynman-Kac model $(M_p^{(\theta,\infty)}, G_p^{(\theta)})$ requires that the samples can be drawn from the Markov transition kernels $M_p^{(\theta,\infty)}$.  However, as discussed at the end of Section \ref{sec:intro-fkm}, in many settings important for real applications, the assumption that the dynamics can be simulated does not hold.

We consider the case where the model $(M_p^{(\theta,\infty)}, G^{(\theta)})$ stems from a discretely and partially observed It\^{o} diffusion process.  Suppose $(X_t')_{t\ge 0}$ solves an It\^{o} stochastic differential equation of the form
$$
\ud X_t' = a^{(\theta)}(X_t') \ud t + b^{(\theta)}(X_t') \ud W_t,
\qquad t\ge 0,
$$
where $\{W_t\}_{t\ge 0}$ is a standard Brownian motion. As in Section \ref{sec:intro-fkm}, we assume that there is some observational process $(Y_t')_{t\ge 0}$, and that observations $\{Y_{t_p}'\}_{p=0}^n$ are obtained at discrete times $\{t_p\}_{p=0}^n$.  With $X_p\defeq X_{t_p}'$ and $Y_p\defeq Y_{t_p}'$, and with $G_p^{(\theta)}(x_p)\defeq g_p^{(\theta)}(y_p| x_p)$, we obtain a model $(M_p^{(\theta,\infty)},G_p^{(\theta)})_{p=0}^n$ which additionally satisfies the SSM conditions (\ref{eq:traditional-ssm-one}-\ref{eq:traditional-ssm-two}).

In some, essentially one-dimensional diffusion settings, where the Lamperti transformation \citep[cf.][]{moeller-madsen} can be applied, $\infty$-inference is possible for $p^{(\theta,\infty)}$ \cite{beskosPR,beskosPRF,fearnheadLRS} and $\pi^{(\infty)}$ \cite{sermaidisPRBF,wang-rao-teh}.
Article \cite{franksJLV} attempts to extend to more settings $\infty$-inference for $p^{(\theta,\infty)}$ and $\pi^{(\infty)}$ in a computationally feasible way. 
The approach of \cite{franksJLV} is based on Euler approximations of the dynamics \citep[cf.][]{kloeden-platen}, multilevel Monte Carlo (MLMC) \cite{heinrich,giles-or},
a particle filter coupling \citep{jasraKLZ}, \emph{debiasing} tricks for MLMC  \cite{mcleish,rhee-glynn}, and an IS type correction \cite{viholaHF}.  We introduce each of these in turn in the following. 

\subsection{Euler approximations}
The Euler approximation amounts to defining a discretisation size $h_\ell \propto 2^{-\ell}$ for $\ell\in \mathbb{N}\cup\{0\}$, and replacing the dynamics of the latent process $(X_t')_{t\ge 0}$ with a discrete-time Markov chain,
$$
X_{t+h_\ell}'
=
X_t' + a^{(\theta)}(X_t') h_\ell + b^{(\theta)}(X_t') ( W_{t+h_\ell} - W_t).
$$
Here, $(W_t)_{t\ge 0}$ is a standard Brownian motion, so that $W_{t+h_\ell} - W_t \sim \mathrm{N}(0, h_\ell)$ is independent of $X_u'$, $u\le t$.

The approximate dynamics corresponds to an approximate transition $M_p^{(\theta,\ell)}$, which, together with the conditionally independent observations, results in a model $(M_p^{(\theta,\ell)}, G_p^{(\theta)})$ satisfying the SSM conditions (\ref{eq:traditional-ssm-one}-\ref{eq:traditional-ssm-two}), with $\ell$-smoother $p^{(\theta,\ell)}(\ud x_{0:n})$ given in \eqref{eq:latent-smoother} in Section \ref{sec:intro-prob}, and with joint $\ell$-posterior $\pi^{(\ell)}(\ud \theta, \ud x_{0:n})$ given in \eqref{eq:ell-posterior}.

Joint $\ell$-inference for $\pi^{(\ell)}$ is possible using PMMH (Algorithm \ref{alg:pmmh}) \cite{andrieuDH}, which has been quite popular in the setting of diffusions \citep[cf.][]{golightly-wilkinson}.  Another $\ell$-inference method \cite{jasraKLZ}, which uses PMMH together with a `multilevel' decomposition, is discussed in the following section.  We reiterate that, in distinction to these methods, the goal in \cite{franksJLV} is to develop a $\infty$-inference method (which is also computationally efficient).
\subsection{Multilevel Monte Carlo}
The idea of MLMC is based on telescoping sums, where each summand is coupled in such a way that leads to lower variance of the resulting estimator \cite{heinrich,giles-or}.
The multilevel decomposition used in \cite{jasraKLZ}, for $\ell_F$-inference in partially observed diffusions,
is based on the telescoping sum in terms of expectations of normalised probabilities,
$$
\pi^{(\ell_F)}(\phi)
=
\sum_{\ell=1}^{\ell_F} \bigg( \pi^{(\ell)}(\phi) - \pi^{(\ell-1)}(\phi)\bigg) + \pi^{(0)}(\phi),  
$$
with $\ell_F\ge 1$ ideally taken quite large.   PMMH chains are run at level $\ell$ and at level $\ell-1$ in each summand, and are coupled to each other using the `approximate coupling' described below.

In \cite{franksJLV}, such a telescoping sum is used not to target an expectation (and where the normalising constants must be simultaneously estimated in each summand), but rather an integral taken with respect to an unnormalised $\ell_F$-smoother,
\begin{equation}\label{eq:unn}
p_u^{(\theta, \ell_F)}(\phi)
=
\sum_{\ell=1}^{\ell_F} \bigg( p_u^{(\theta,\ell)}(\phi) - p_u^{(\theta,\ell-1)}(\phi)\bigg) + p_u^{(\theta,0)}(\phi),  
\end{equation}
with $\ell_F\ge 1$ ideally taken quite large.  The quality of the approximation as measured by the variance depends on the coupling used for each increment
$$
p_u^{(\theta,\ell)}(\phi) - p_u^{(\theta,\ell-1)}(\phi).
$$

The algorithm used in \cite{franksJLV} to unbiasedly estimate this difference is given in Algorithm \ref{alg:delta-pf}, which we refer to as the `delta PF' ($\Delta$PF).
The coupling used is the `approximate coupling' of \cite{jasraKLZ}.
This coupling is based on a change of measure of the Feynman-Kac model on a joint path space, which, together with an importance sampling correction of the particle filter output, leads to the $\Delta$PF.

\subsection{Coupling of Feynman-Kac models}
Suppose $(M_p^{(\theta,\ell)}, G_p^{(\theta)})_{p=0}^n$ and $(M_p^{(\theta,\ell-1)}, G_p^{(\theta)})_{p=0}^n$ are two Feynman-Kac models.  We describe a coupling of them as follows.  For some fixed $\ell\ge 1$,  $\check{M}_p^{(\theta,\ell)}(\check{x}_{0:p-1}, \ud \check{x}_p)$ is assumed to be a coupling of the $\ell$ and $\ell-1$ level transitions, that is,
\begin{align*}
 \check{M}_p^{(\theta,\ell)} (\check{x}_{0:p-1}, A\times\X) &= M_p^{(\theta,\ell)}( x_{0:p-1}^{(\ell)}, A),
\\
\check{M}_p^{(\theta,\ell-1)} (\check{x}_{0:p-1}, \X\times A) &= M_p^{(\theta,\ell-1)}( x_{0:p-1}^{(\ell-1)}, A),
\end{align*}
for $A\in \mathcal{B}(\X)$ and with the notation $\check{x}_{0:p}=(x_{0:p}^{(\ell)}, x_{0:p}^{(\ell-1)})$ denoting an element in the space $\X^{2(p+1)}$, and we set 
\begin{equation}\label{eq:coupled-potential}
\check{G}_{0:p}^{(\theta)}(\check{x}_{0:p})
=
\frac{1}{2}\bigg(
G_p^{(\theta)}( x_{0:p}^{(\ell)}) + G_p^{(\theta)}( x_{0:p}^{(\ell-1)}) \bigg).
\end{equation}
The dynamics $\check{M}_p^{(\theta,\ell)}$ is typically obtained in the diffusion context by using a common Brownian path for mesh discretisation levels $\ell$ and $\ell-1$.  Other choices for $\check{G}_p^{(\theta)}$ are possible then the choice \eqref{eq:coupled-potential} used in \cite{franksJLV}.  The important point is that $\check{G}_p^{(\theta)}(\check{x}_{0:p})>0$ whenever $G_p^{(\theta)}(x_{0:p}^{(\ell)})>0$ or $G_p^{(\theta)}(x_{0:p}^{(\ell-1)})>0$.  This ensures that the estimator $\Delta^{(\theta,\ell)}(\phi)$ from the $\Delta$PF (Algorithm \ref{alg:delta-pf}) is unbiased \citep[][Prop.~3]{franksJLV}: for bounded $\phi:\X^{n+1}\rightarrow \R$,
$$
\E[ \Delta^{(\theta,\ell)}(\phi)]
=
p_u^{(\theta,\ell)}(\phi) - p_u^{(\theta,\ell-1)}(\phi).
$$
\begin{algorithm}
  \caption{Delta particle filter ($\Delta$PF) for $(\check{M}_p^{(\theta,\ell)}, \check{G}_p^{(\theta)})_{p=0}^n$, with $\ell \ge 1$ and with $N\ge 1$ particles.}
  \label{alg:delta-pf}
  \begin{enumerate}[(i)]
  \item  Run PF (Algorithm \ref{alg:pf}) for $(\check{M}_p^{(\theta,\ell)}, \check{G}_p^{(\theta)})_{p=0}^n$,
    outputting $( V_n^{(i)},\check{X}_{0:n}^{(i)})_{i=1}^N$.
  \item Output $\Delta^{(\theta,\ell)}$, where, for $\phi:\X^{n+1}\to\R$, 
    \begin{equation}\label{eq:delta-pf-estimator}\nonumber
    \Delta^{(\theta,\ell)}(\phi)\defeq \sum_{i=1}^N V_n^{(i)}\bigg( \overline{w}_\ell( \check{X}_{0:n}^{(i)}) \phi(X^{(\ell,i)}_{0:n}) - \underbar{w}_\ell(\check{X}^{(i)}_{0:n}) \phi(X^{(\ell-1,i)}_{0:n})\bigg)
    \end{equation}
    where
    $$
    \overline{w}_\ell(\check{X}_{0:n}) \defeq \frac{\prod_{p=0}^n G_p^{(\theta)}(X_{0:p}^{(\ell)})}{\prod_{p=0}^n \check{G}_p^{(\theta)}(\check{X}_{0:p})}
    \qquad
    \text{and}
    \qquad
    \underbar{w}_\ell(\check{X}_{0:n}) \defeq \frac{\prod_{p=0}^n G_p^{(\theta)}(X_{0:p}^{(\ell-1)})}{\prod_{p=0}^n \check{G}_p^{(\theta)}(\check{X}_{0:p})}.
    $$
\end{enumerate}
    \end{algorithm}
We can then estimate $p_u^{(\theta,\ell_F)}$  unbiasedly using MLMC.  Namely,
\begin{equation}\label{eq:det-ell}
  \E\big[
\mulp_{m_{0}}^{(\theta,0)}(\phi) +
    \mulp_{m_{1:F}}^{(\theta,1:\ell_F)}(\phi) \big]= p_u^{(\theta,\ell_F)}(\phi),
\end{equation}
where
\begin{equation*}
  \mulp_{m_{0}}^{(\theta,0)}(\phi)
  \defeq
  \frac{1}{m_0} \sum_{i=1}^{m_0} \hat{p}_{u,i}^{(\theta,0)}(\phi),
\end{equation*}
with $\{\hat{p}_{u,i}^{(\theta,0)} (\phi) \}_{i=1}^{m_0}$ independently run versions of the estimator $\hat{p}_u^{(\theta,0)}(\phi) = \sum_{i=1}^N V^{(i)} \phi(\mathbf{X}^{(i)})$ from the output $(V^{(1:N)}, \mathbf{X}^{(1:N)})$ of the PF (Algorithm \ref{alg:pf}) run for the model $(M_p^{(\theta,0)}, G_p^{(\theta)})_{p=0}^n$, and where
  $$
  \mulp_{m_{1:F}}^{(\theta,1:\ell_F)}(\phi)
  \defeq
  \sum_{\ell=1}^{\ell_F} \frac{1}{m_\ell} \sum_{i=1}^{m_\ell} \Delta_i^{(\theta,\ell)}(\phi),
$$
with
$\{\Delta_i^{(\theta,\ell)}(\phi)\}_{i=1}^{m_\ell}$ estimators formed from independent runs of the $\Delta$PF (Algorithm \ref{alg:delta-pf}) run for the model $(\check{M}_p^{(\theta,\ell)}, \check{G}_p^{(\theta)})_{p=0}^n$.

The approach based on \eqref{eq:det-ell} allows for efficient MLMC estimation of the unnormalised  $\ell_F$-smoother $p_u^{(\theta,l_F)}$, over the latent states.  If we were content with  joint $\ell_F$-inference, then we could apply already the IS type correction of MCMC as in Algorithm \ref{alg:is},
with regularised `likelihood' estimate
$$
L(\Theta_k) \defeq   \mulp_{m_0}^{(\Theta_k,0)}(1)+\epsilon
$$
in the acceptance ratio \eqref{eq:acc-da-one},
with $\epsilon \ge 0$, and
with IS weights 
$$
\xi_k(\phi) \defeq
\frac{ \mulp_{m_0}^{(\Theta_k,0)}(\phi) + \mulp_{m_{1:F}}^{(\Theta_k,1:\ell_F)}(\phi)}{ \mulp_{m_0}^{(\Theta_k,0)}(1)+\epsilon},
 $$
which are allowed to take negative values \citep[cf.][]{viholaHF}.
This would provide an efficient MLMC alternative method to the PMMH or the algorithm in \cite{jasraKLZ} for inference with respect to $\pi^{(l_F)}$.  Instead, we wish to go one (infinite!) step further, and target $\pi^{(\infty)}$.

\subsection{Debiasing techniques}\label{sec:diffusions-debiasing}
Debiased MLMC \cite{mcleish,rhee-glynn,vihola-unbiased} is based on randomising the running level used in deterministic MLMC (with a reweighting), as follows.

We assume that $(p_\ell)_{\ell \ge 1}$ is a probability mass function (p.m.f.) on $\mathbb{N}$ satisfying $p_\ell >0$ for all $\ell\ge 1$.  We also assume that
$$
p_u^{(\theta,\ell)}(\phi) \xrightarrow{\ell\to\infty} p_u^{(\theta,\infty)}(\phi),
$$
for all bounded $\phi:\X^{n+1}\rightarrow \R$,
which is not too difficult to verify in our setting under certain boundedness assumptions, because of the known convergence properties of the Euler approximation \citep[cf.][]{kloeden-platen}. 
With $L\sim (p_\ell)$, the \emph{single-term debiased MLMC estimator} of \cite{rhee-glynn} in our case is given by $p_L^{-1} \Delta^{(\theta,L)}(\phi)$, which satisfies
$$
\E[ p_L^{-1} \Delta^{(\theta,L)}(\phi) ]
=
p_u^{(\theta,\infty)}(\phi) - p_u^{(\theta,0)}(\phi).
$$

Adding an independent `zeroth level' estimate $\hat{p}_u^{(\theta,0)}(\phi)  \defeq \sum_{i=1}^N V^{(i)} \phi(\mathbf{X}^{(i)})$, formed from the output $(V^{(i)},\mathbf{X}^{(i)})_{i=1}^N$ of PF (Algorithm \ref{alg:pf}) run for the model $(M_p^{(\theta,0)}, G_p^{(\theta)})_{p=0}^n$, we set
\begin{equation}\label{eq:tilde-delta}
\tilde{\Delta}^{(\theta)}(\phi) \defeq \frac{1}{p_L} \Delta^{(\theta,L)}(\phi) + \hat{p}_u^{(\theta,0)}(\phi),
\end{equation}
to obtain that
$$
\E[ \tilde{\Delta}^{(\theta)} (\phi)]
=
p_u^{(\theta,\infty)}(\phi).
$$
By using a self-normalised estimator to take care of the normalising constant, this already allows for consistent inference over the latents.  That is,
as in \eqref{eq:ratio-estimator} of Section \ref{sec:intro-pf}, if $\{\tilde{\Delta}_k^{(\theta)}\}_{k=1}^m$ for $m\ge 1$ are independently run to form estimator functionals of the form \eqref{eq:tilde-delta}, then
$$
\frac{\sum_{k=1}^m \tilde{\Delta}_k^{(\theta)}(\phi)}{\sum_{k=1}^m \tilde{\Delta}_k^{(\theta)}(1)}\longrightarrow p^{(\theta,\infty)}(\phi),
\qquad
\text{almost surely,}
$$
as $m\to\infty$ \citep[Prop.~7]{franksJLV}.
\subsection{Joint inference using importance sampling type correction}
Recall that our original goal was joint $\infty$-inference (for $\pi^{(\infty)}$).  To do this, we will use Algorithm \ref{alg:is-diffusion}, which is similar to Algorithm \ref{alg:is}, but which uses a multilevel IS type correction based on the randomised $\Delta$PF output.  Consistency was also detailed in \cite{viholaHF} for IS type correction involving negative weights as in Algorithm \ref{alg:is-diffusion}, which can occur frequently in the multilevel context which we consider here.

\begin{algorithm}[]
  \caption{MCMC-IS for joint $\infty$-inference for diffusions based on debiased IS type correction, with $m\ge 1$, $\epsilon \ge 0$, p.m.f.~$(p_\ell)$  on $\N$, and $N_\ell \ge 1$ for all $\ell \ge 0$.}
  \label{alg:is-diffusion}
  \begin{enumerate}[(P1)]
  \item\label{alg:is-diffusion-one}
        With $(\Theta_0, V_0^{(i)}, \mathbf{X}_0^{(i)})_{i=1}^{N_0}$ given, for $k=1,\ldots, m$, do:
\begin{enumerate}[(i)]
  \item
Sample $\Theta'\sim q(\uarg| \Theta_{k-1})$ from a transition kernel $q$.  
  \item 
    Run PF (Algorithm \ref{alg:pf}) for $(M_p^{(\Theta',0)}, G_p^{(\Theta')})$, with output $(V^{'(i)}, \mathbf{X}^{'(i)})_{i=1}^{N_0}$.
\item
    Accept, setting $(\Theta_k,V_k^{(i)},\mathbf{X}_k^{(i)})_{i=1}^{N_0}\leftarrow (\Theta', {V'}^{(i)}, {\mathbf{X}'}^{(i)})_{i=1}^{N_0}$, with probability 
    $$
    \min\bigg\{
1, \frac{\pr(\Theta') \big(\epsilon + \sum_{i=1}^{N_0} {V'}^{(i)} \big)  q(\Theta_{k-1}| \Theta')}{\pr(\Theta_{k-1}) \big( \epsilon + \sum_{i=1}^{N_0} V_{k-1}^{(i)} \big) q(\Theta'| \Theta_{k-1})}
\bigg\}.
$$
Otherwise, reject, setting $(\Theta_k, V_k^{(i)}, \mathbf{X}_k^{(i)})_{i=1}^{N_0}\leftarrow (\Theta_{k-1}, V_{k-1}^{(i)}, \mathbf{X}_{k-1}^{(i)})_{i=1}^{N_0}$.
\end{enumerate}
\item\label{alg:is-diffusion-two}  
For all $k\in \{1{:}m\}$,
  \begin{enumerate}[(i)]
\item  Sample $L_k\sim (p_\ell)$.
\item  Run $\Delta$PF (Algorithm \ref{alg:delta-pf}) for $(\check{M}^{(\Theta_k, L_k)}, \check{G}^{(\Theta_k)})$ with $N_{L_k}$ particles, outputting $\Delta^{(\Theta_k,L_k)}$.
  \newline
  With
  $
  \xi_k(\phi)
  $
  defined for $\phi:\X^{n+1}\rightarrow \R$, form the estimator
  $$
  E_m^{IS}(f)\defeq
  \frac{\sum_{k=1}^m \xi_k(f^{(\Theta_k)})}{\sum_{k=1}^m \xi_k(1)},
  \quad
    \xi_k(\phi)
  \defeq
  \frac{ p_{L_k}^{-1} \Delta^{(\Theta_k,L_k)}(\phi) + \sum_{i=1}^{N_0} V_k^{(i)} \phi(\mathbf{X}_k^{(i)})}{\epsilon + \sum_{i=1}^{N_0} V_k^{(i)}}. 
  $$
\end{enumerate}
\end{enumerate}
\end{algorithm}
The likelihood support condition \eqref{eq:support-condition} mentioned for Algorithm \ref{alg:is} can be achieved by using $\epsilon>0$.\footnote{It is closely linked to `defensive importance sampling' \cite{hesterberg}, but its optimal choice in terms of efficiency is not known.}  We also need finiteness of the variance of the randomised $\Delta$PF, $p_L^{-1} \Delta^{(\theta,L)}$, in order to guarantee that the debiased MLMC works correctly \citep[cf.][]{rhee-glynn}, and that the MCMC-IS (Algorithm \ref{alg:is-diffusion}) can have finite asymptotic variance \citep[cf.][Prof.~13]{franksJLV}. That is, we need to show that
\begin{equation}\label{eq:delta-pf-var}
\var \bigg( \frac{1}{p_L} \Delta^{(\theta,L)}(\phi) \bigg)
=
\sum_{\ell\ge 1} \frac{\E\big[ (\Delta^{(\theta,\ell)}(\phi))^2\big]}{p_\ell}
- \Big(p_u^{(\theta,\infty)}(\phi) - p_u^{(\theta,0)}(\phi)\Big)^2
\end{equation}
is finite, uniformly in $\theta\in \T$.  This requires showing that the variance of $\Delta^{(\theta,\ell)}(\phi)$ decays at a sufficient rate relative to $p_\ell$ as $\ell$ increases.

Under some standard (stringent) assumptions used elsewhere in the literature, the results of the technical analysis are formulated in \citep[Cor.~$9$]{franksJLV}.  In the case of standard Euler approximation, the result says that
\begin{equation}\label{eq:delta-pf-cor}
\E\big[ (\Delta^{(\theta,\ell)}(\phi))^2\big]
\le
C\Big( \frac{2^{-\ell}}{N_\ell} + 2^{-2 \ell}\Big),
\end{equation}
where $C>0$ is a constant which does not depend on $N_\ell\ge 1$, $\ell\ge 1$, or $\theta\in\T$, where $N_\ell$ particles are used in the $\Delta$PF run at level $\ell$.  Hence, with $N_\ell=N$ constant, by taking $p_\ell \propto 2^{-r \ell}$, with $r<1$, \eqref{eq:delta-pf-var} will be finite.  More generally, if $N_\ell\propto 2^{\rho \ell}$ with $\rho\in [0,1]$, then we see that we can take $p_\ell \propto 2^{-r\ell}$ with $r<1 + \rho$, so that \eqref{eq:delta-pf-var} will be finite.

The assumptions needed to prove the bound \eqref{eq:delta-pf-cor} in \cite{franksJLV} are on the diffusion \citep[e.g.][]{kloeden-platen}, in terms of uniform ellipticity and globally Lipschitz diffusion terms, as well as on the Feynman-Kac model \citep[e.g.][]{del-moral}, in terms of globally Lipschitz potentials and transitions and lower and upper bounded potentials. The results of the analysis are based on a global error martingale decomposition \citep[cf.][]{del-moral} in terms of the local sampling error of the particle filter run for the coupled Feynman-Kac model, and on an analysis of the $\Delta$PF in the diffusion context.    

\subsection{Computational efficiency and allocations}\label{sec:diffusions-efficiency}
We have seen that under some assumptions, the finiteness of the variance of the randomised $\Delta$PF can be verified for any $N_\ell\ge 1$ and for sufficiently heavy-tailed $(p_\ell)$ \citep[Cor.~9]{franksJLV}.  However, the use of a heavy-tailed p.m.f.~$(p_\ell)$ can lead to excessive use of computational resources, and we must therefore try to use thinner-tailed p.m.f.s $(p_\ell)$ and optimal number of particles $N_\ell$ at level $\ell$ in order to minimise the inverse relative efficiency (IRE) \cite{glynn-whitt} which measures the computational cost.

Let $(\Theta_k)_{k\ge 1}$ be the marginal Markov chain of Algorithm \ref{alg:is-diffusion}, and $L_k\sim (p_\ell)$ for $k\ge 1$.  
With terminology similar to \cite{glynn-whitt}, who consider the i.i.d.\footnote{independent and identically distributed} case for $(\tau_k)_{k\ge 1}$, we assume that the \emph{total computational cost} to run Algorithm \ref{alg:is-diffusion} for $m$ iterations is
\begin{equation}\label{eq:total-computational-cost}\nonumber
\mathscr{C}(m)
\defeq
\sum_{k=1}^m
\tau_{k},
\end{equation}
where $(\tau_k)_{k\ge 1}$ are conditionally independent positive random variables given $(\Theta_k, L_k)_{k\ge 1}$, where $\tau_k$  depends only on $\Theta_k$ and $L_k$.  Given some \emph{budget} $\kappa\in \mathbb{R}_{\ge 0}$, the \emph{realised length} of the chain is
\begin{equation}\label{eq:realised-length}\nonumber
\mathscr{M}(\kappa)
\defeq
\max\big\{ m\in \mathbb{N}_{\ge 0} | \mathscr{C}(m) \le \kappa\big\}.
\end{equation}
Then, if for some number $\tau>0$, 
$$
\frac{1}{m} \sum_{k=1}^m \tau_k \xrightarrow{m\to \infty} \tau,
\qquad
\text{almost surely,}
$$
and if the MCMC-IS estimator satisfies a CLT with asymptotic variance $\sigma^2(f)$, then \cite{glynn-whitt,franksJLV}
$$
\sqrt{\kappa}\big[
  E_{\mathscr{M}(\kappa)}^{IS}(f) - \pi^{(\infty)}(f)
  \big]
\xrightarrow{\kappa\to\infty}
\mathrm{N}\big(0, \tau \sigma^2(f)\big),
\qquad
\text{in distribution},
$$
 and $\tau\sigma^2(f)$ is the IRE.  We thus extend the discussion of \cite{glynn-whitt} to non-i.i.d.~$(\tau_k)_{k\ge 1}$.  

 Using this computational efficiency framework, similar to \cite{rhee-glynn} who consider the i.i.d.~case in traditional MLMC, it is possible to consider the matter of computational complexity and optimal allocation of resources in Algorithm \ref{alg:is-diffusion}.  Suppose a CLT holds for the MCMC-IS estimator of Algorithm \ref{alg:is-diffusion} with finite asymptotic variance \citep[cf.][Prop.~13]{franksJLV}.  Let $\epsilon>0$ and $\delta\in(0,1)$ be given.  In order to have
 $$
\P[ \abs{E_m^{IS}(f) - \pi^{(\infty)}(f)}\le \epsilon] \ge 1-\delta,
 $$
by the Chebyshev inequality and using the standard $m^{-1}$ mean squared error convergence rate for MCMC, we need that $m$ is of order $\epsilon^{-2}$, denoted  $m=O(\epsilon^{-2})$.\footnote{That is, with $m=m(\epsilon)$, we have $m(\epsilon)/\epsilon^{-2}\to C$ as $\epsilon\to 0$, some $C>0$.}   The question is then how we can minimise the computational complexity given by $\mathscr{C}(m)$ when $m=O(\epsilon^{-2})$, by adjusting $p_\ell$ and $ N_\ell$, while keeping the variance \eqref{eq:delta-pf-var} finite.\footnote{Besides for the debiasing \cite{rhee-glynn} to work, the asymptotic variance \citep[see][Prop.~13]{franksJLV} of the MCMC-IS estimator of Algorithm \ref{alg:is-diffusion} has a part from the marginal MCMC, as well as from the IS type correction.  The latter is finite if the variance \eqref{eq:delta-pf-var} is finite.}    Assuming
$$
\E[ \tau_{k}| \Theta_k=\theta,\,L_k=\ell] \le C 2^{\ell(1+\rho)},
$$
where $C$ does not depend on $\theta$ or $\ell$, then it is shown in \citep[Prop.~24]{franksJLV} that for all $q>2$, $\eta>1$, the computational cost
\begin{equation}\label{eq:cost-order-q}
O\big(\epsilon^{-2} \abs{\log_2 \epsilon}^q\big)
\end{equation}
can be obtained for sufficiently small $\epsilon$, if $p_\ell$ and $N_\ell$ are chosen to be
\begin{equation}\label{eq:p-ell-log}
  p_\ell \propto 2^{-\ell(1+\rho)} \ell [\log_2(\ell +1)]^\eta
  \qquad\text{and}\qquad
  N_\ell \propto 2^{\rho \ell}
\end{equation}
for $\rho \in [0,1]$.  This choice for $p_\ell$ and $N_\ell$ ensures that the variance \eqref{eq:delta-pf-var} is finite, and suggests\footnote{by disregarding the factor $\ell [\log_2(\ell+1)]^\eta$ in \eqref{eq:p-ell-log}} the choice
\begin{equation}\label{eq:p-ell-suggestion}
p_\ell \propto 2^{-\ell(1+\rho)}
\qquad\text{and}\qquad
N_\ell \propto 2^{\rho\ell}
\end{equation}
for $\rho\in [0,1]$.  The computational cost \eqref{eq:cost-order-q} is the same as that of \citep[Prop.~4]{rhee-glynn} for the single-term estimator in traditional, randomised MLMC.  It is also  very close to
$$
O\big(\epsilon^{-2} (\log_2 \epsilon)^2\big),
$$
(recall $q>2$), which is the well-known computational complexity order  \cite{giles-or} in the traditional,  deterministic MLMC.

 The result \eqref{eq:p-ell-log} shows that in case of Euler approximation, there is in fact a parametrisation of recommended choices for particle number $N_\ell$ and p.m.f.~$(p_\ell)$, all of which share the same order of computational complexity to obtain a given precision, under certain assumptions  such as previously explained for $\tau_k$.
Then \eqref{eq:p-ell-log} (or the simplified suggestion \eqref{eq:p-ell-suggestion}) should lead to a proper usage of computational resources, in order to keep both the asymptotic variance and the total cost jointly small, and therefore the IRE small.  The \emph{order} of computational complexity is the same along the parametrisation in terms of $\rho\in [0,1]$, but it is still unknown whether a certain choice of $\rho$ will usually lead to the best choice for $N_\ell$ and $(p_\ell)$.  In an experiment in \cite{franksJLV} concerning a geometric Brownian motion, the choice $\rho=0$ performed better than the choice $\rho=1$ in the allocation \eqref{eq:p-ell-log}.   We leave, for now, the optimal choice of $\rho$ for future research and experiment.

\section{Inference via approximate Bayesian computation}
\label{sec:abc}

We assume a Bayesian model as in Section \ref{sec:intro-bayesian}, with fixed observation denoted $y^*\in\Y$, prior $\pr(\theta)$, and likelihood $L(\theta)=p^{(\theta)}( y^*)$, which is assumed to be intractable.  Although the data distribution $p^{(\theta)}(\uarg)$ can not be evaluated, we assume that it is possible to sample data $y\sim p^{(\theta)}(\uarg)$ from it.  Let $d(y,y')$ be a pseudo-metric\footnote{
  That is, for all $y_1, y_2, y_3\in \Y$, it holds $d(y_1,y_2)\ge 0$, $d(y_1,y_2) = d(y_2,y_1)$, and $d(y_1,y_3) \le d(y_1,y_2)+ d(y_2,y_3)$. For example, $d(y_1,y_2) = \norm{s(y_1) - s(y_2)}$ where $s:\Y\to \R^{n_y}$ is some (summary) statistic \citep[cf.][]{prangle-summary}.}
 on $\Y^2$.   With \emph{tolerance} $\epsilon>0$, we then define
$$
p_u^{(\theta,1/\epsilon)}(\ud y)
\defeq
p^{(\theta)}(\ud y) \mathbf{1}\big( d(y,y^*) \le \epsilon \big),
\footnote{The quantity `$1/\epsilon$' can be thought of as denoting the level of `precision.'}  
$$
Approximate Bayesian computation (ABC) (see \citep[][]{sissonFB} for a review) is based on using the family
$
\mathscr{P}_{1/ \epsilon}
\defeq
\{ p^{(\theta,1/\epsilon)} \}_{\theta\in\T}
$
of approximate probabilities, where
\begin{equation}\label{eq:abc-data-dist}\nonumber
p^{(\theta,1/ \epsilon)}(\ud y)
\defeq
\frac{p_u^{(\theta,1/ \epsilon)}(\ud y)}{L^{(1/\epsilon)}(\theta)},
\end{equation}
with ABC likelihood,
\begin{equation}\label{eq:abc-likelihood}\nonumber
L^{(1/ \epsilon)}(\theta)
\defeq
\int p^{(\theta)}(\ud y) \mathbf{1}\big( d(y,y^*) \le \epsilon\big).
\end{equation}
Then $\mathscr{P}_{1/ \epsilon}$ become families of increasingly `better' approximations as $\epsilon$ goes to $0$.  However, it is important to keep in mind that it is only approximate even in the limit, since in general,
$$
L^{(\infty)}(\theta)
\defeq 
\lim_{\epsilon\to 0} L^{(1/ \epsilon)}(\theta)
\neq L(\theta).
\footnote{This is in general the case.  However, there can be equality if, for example, $d(y,y')$ is a metric, or, in particular, a metric formed from composition of a sufficient statistic with a Euclidean norm $\norm{\uarg}$ \citep[cf.][]{prangle-summary}.}
$$
The  ABC posterior is then given by
$$
\pi^{(1/ \epsilon)}(\theta) \propto \pr(\theta) L^{(1/\epsilon)}(\theta).
$$
A method of inference for the ABC posterior which we consider is the  ABC-MCMC (Algorithm \ref{alg:abc-mcmc}), as suggested by  \cite{marjoramMPT}, which may also be viewed as a pseudo-marginal MCMC \cite{andrieu-roberts}, with
$$
\E_\theta\big[\charfun{d (Y,y^*)\le \epsilon}\big] =L^{(1/\epsilon)}(\theta).
$$

\subsection{Choosing the tolerance in ABC-MCMC}\label{sec:abc-choice}
The choice of tolerance $\eps$ is a difficult question in ABC-MCMC \citep[cf.][]{sisson-fan}.  Namely, a large choice of $\epsilon$  leads to large bias, but to computational inefficiency if $\eps$ is small. To see this, note that if $\eps$ is small, then a proposed state is hardly ever accepted, since $\charfun{d(Y_k',y^*)\le \eps}$ is usually $0$.  If $\eps$ is large, then $L^{(1/\epsilon)}(\theta)\approx 1$ is nearly constant in $\theta$ and so ABC-MCMC is essentially targeting the prior model, which is uninformative for Bayesian posterior inference.

Article \cite{vihola-franks} attempts to deal with the issue of tolerance choice in ABC-MCMC, by using an inflated and adaptively tuned  tolerance parameter in order to maximise efficiency of the MCMC, and then to use a post-correction, importance sampling step, to remove bias \cite{wegmannLE} as well as to quantify uncertainty with proposed approximate confidence intervals.  

The tolerance adaptive ABC-MCMC (Algorithm \ref{alg:ta}), which is run during burn-in for some number of iterations $n_b$, is an adaptive MCMC \citep[cf.][]{andrieu-thoms} targeting a user-specified overall acceptance probability $\alpha^*\in (0,1)$.
In experiments in \cite{vihola-franks}, a value of $\alpha^*=10\%$ was used, which ensures sufficient mixing and number of different samples from the MCMC.  
We provide convergence theorems in \cite{vihola-franks} for the adaptive algorithm under two sets of assumptions.  The simpler set of assumptions essentially requires that the proposal $q(\theta'|\theta)>0$ is uniformly bounded away from zero, and $\epsilon_k$ is bounded away from zero for all $k\ge 1$  almost surely.  The former assumption on $q$ is removed in the more general set of assumptions.  Removing the assumption on $\epsilon_k$ might be possible, based on projections \citep[cf.][]{andrieuMP}.

\begin{algorithm}
  \caption{ABC-MCMC($\eps$).  Given $\Theta_0\in \T$ with $\pr(\Theta_0)>0$, run the following for $k=0,\ldots n-1$:}
\label{alg:abc-mcmc}
    \begin{enumerate}[(i)]
    \item Sample $\Theta_k' \sim q( \uarg| \Theta_k)$.
    \item Sample $Y_k' \sim p^{(\Theta_k')}(\uarg)$.
      \item Accept, setting $(\Theta_{k+1},Y_{k+1})\leftarrow(\Theta_k',Y_k')$, with probability $\alpha_{\eps}(\Theta_k, \Theta_k',Y_k')$,  where
        \begin{equation}
          \label{eq:acc-epsilon}
        \alpha_{\eps}(\theta, \theta',y')
        \defeq
        \min\bigg\{1, \frac{\pr(\theta') q(\theta| \theta')}{\pr(\theta) q(\theta'|\theta)}\bigg\}
        \charfun{d(y',y^*)\le \eps}.
      \end{equation}
      Else, reject, by setting $(\Theta_{k+1}, Y_{k+1}) \leftarrow (\Theta_{k}, Y_{k})$.
      \end{enumerate}
\end{algorithm}

\subsection{Approximate confidence intervals}\label{sec:abc-ci}

An approximate estimator for the asymptotic variance of the post-corrected ABC-MCMC has been suggested in \citep[][Alg.~6]{vihola-franks}, which can be used for the construction of (approximate) confidence intervals.

Suppose that $\hat{\tau}_{\epsilon_0}(f)$ is an estimate for the integrated auto-correlation time for ABC-MCMC($\epsilon_0$), 
\begin{equation}\label{eq:iact}
\tau_{\epsilon_0}(f)\defeq
\sum_{k\ge 1} \mathrm{Corr}\big(f(\vartheta_0),f(\vartheta_k)\big),
\qquad
\vartheta_0\sim \pi^{(1/\epsilon_0)}(\uarg),
\end{equation}
perhaps using a windowed sample auto-correlation estimator \citep[cf.][]{harris-windows}.   
Also define the following estimator for the function variance,
\begin{equation}\label{eq:variance-est}
    S_{\epsilon_{0}, \epsilon}(f)
    \defeq
     \sum_{k=1}^n \frac{\mathbf{1}\big(d(Y_k,y^*) \le \epsilon\big) \big( f(\Theta_k) - E_{\epsilon_{0},\epsilon}(f)\big)^2}{\big( \sum_{j=1}^n \mathbf{1}\big( d(Y_j,y^*) \le \epsilon\big)\, \big)^{2} }.
\end{equation}
     The approximate confidence interval then takes the form
     $$
\Big[ E_{\epsilon_0, \epsilon}(f) \pm \beta \sqrt{\hat{\tau}_{\epsilon_0}(f) S_{\epsilon_0,\epsilon}(f) } \Big],
     $$
where $\beta>0$ corresponds to the standard normal quantile.

We remark that there is some theoretical backing for the  approximate confidence interval, based on an exact formula for the integrated auto-correlation time of the post-corrected chain \citep[Thm.~7]{vihola-franks}. The relevance of the approximate confidence intervals is also verified in some experiments in \cite{vihola-franks}.

\begin{algorithm}
  \caption{
    TA($n_b$).  Given
  $\Theta_0\in \T$ with $\pr(\Theta_0)>0$, $\eps_0\defeq d(Y_0,y^*)>0$ with $Y_0\in\Y$, $\alpha^*=.1$, and  step sizes $\gamma_k=k^{-2/3}$.}
\label{alg:ta}
\begin{flushleft}
    For $k=0,\ldots, n_b-1$,
  \end{flushleft}
\begin{enumerate}[(i)]
    \item Sample $\Theta_k' \sim q( \uarg| \Theta_k)$.
    \item Sample $Y_k'\sim p^{(\Theta_k')}(\uarg)$.
      \item Accept, setting $(\Theta_{k+1},Y_{k+1})\leftarrow(\Theta_k',Y_k')$, with probability $\alpha_{\eps_k}(\Theta_k,\Theta_k', Y_k')$, with $\alpha_\epsilon$ defined in \eqref{eq:acc-epsilon}.
Otherwise, reject, setting $(\Theta_{k+1}, Y_{k+1}) \leftarrow (\Theta_{k}, Y_{k})$.  
\item \label{alg:ta-adaptation}
  $\log \eps_{k+1} \leftarrow \log \eps_k + \gamma_k\big( \alpha^* - \alpha_{\eps_k}(\Theta_k,\Theta_k',Y_k')\big)$.
    \end{enumerate}
\begin{flushleft}
 Output $(\Theta_{n_b},\eps_{n_b})$.
  \end{flushleft}
\end{algorithm}

\subsection{Adaptive ABC-MCMC with post-correction}
The approach of \cite{vihola-franks} then takes the form of Algorithm \ref{alg:abc-is}.  In regards to the adaptive ABC-MCMC, also the proposal covariance matrix $q$ is best updated as in \cite{haarioST,andrieu-moulines}.
The estimator $E_{\eps_0,\eps}(f)$ can be calculated effortlessly for all $\eps\in (0,\eps_0]$ by sorting beforehand the samples $\Theta_k$ according to their corresponding distances $T_k$.
  
  In experiments in \cite{vihola-franks}, for example in a Lotka-Volterra model involving two reagents and three reactions \citep[cf.][]{boysWK}, it was found that Algorithm \ref{alg:abc-is} delivers a robust approach to inference in ABC models.  In particular, the post-processing estimators were found to be competitive with direct ABC-MCMC with pre-tuned tolerance and starting value, the approximate confidence interval provided good coverage, and the adaptive ABC-MCMC allowed for essentially arbitrary initial choice of tolerance and starting value from the prior.

  Compared to direct ABC-MCMC($\epsilon$), the approach based on slightly inflated tolerance and post-correction, was shown to be competitive in experiments in \cite{vihola-franks}.
  An upper bound for the asymptotic variance of the ABC-MCMC($\epsilon_0$)  with post-correction to $\epsilon$, in terms of that of a direct ABC-MCMC($\epsilon$), is given in \citep[Thm.~8]{vihola-franks}.  It is a direct application of the Peskun type ordering for importance sampling \cite{franks-vihola} stated previously in \eqref{eq:covariance-is-one}, where the upper bound guarantee becomes an equality as $\abs{\epsilon_0 - \epsilon}\to 0$.      
  
\begin{algorithm}
\caption{Given $n_b,\,n\ge1$ perform the following:}
\label{alg:abc-is}
\begin{enumerate}[(i)]
\item Run TA($n_b$) (Algorithm \ref{alg:ta}), and call the output $(\Theta_0,\eps_0)$.
\item Run ABC-MCMC($\eps_0$) (Algorithm \ref{alg:abc-mcmc}) for $n$ iterations, with starting values $(\Theta_0, \eps_0)$, outputting $(\Theta_k,Y_k)_{k=1}^n$.
\item\label{alg:abc-is-is} For all $\eps\le \eps_0$, an estimator for $\pi^{(1/\eps)}(f)$ is given by
  $$
  E_{\eps_0,\eps}(f)
  \defeq
  \frac{\sum_{k=1}^n \charfun{d(Y_k,y^*)\le \eps} f(\Theta_k)}{\sum_{k=1}^n \charfun{d(Y_k,y^*)\le \eps}}.
  $$
\item With $\hat{\tau}_{\epsilon}(f)$ an estimate of \eqref{eq:iact}, $S_{\epsilon_0,\epsilon}(f)$ calculated as in \eqref{eq:variance-est}, and $\beta>0$ corresponding to the desired standard normal quantile, report the approximate confidence interval
  $$
  \Big[ E_{\epsilon_0, \epsilon}(f) \pm \beta \sqrt{\hat{\tau}_{\epsilon_0}(f) S_{\epsilon_0,\epsilon}(f) } \Big].
  $$
  \end{enumerate}
  \end{algorithm}

\subsection{Convergence of the tolerance adaptive ABC-MCMC}
We briefly discuss the general approach to the convergence proofs of the tolerance adaptive ABC-MCMC.  To obtain a setup fitting within the framework of stochastic approximation \citep[cf.][]{andrieu-moulines,andrieuMP}, we write the tolerance adaptation update in Algorithm \ref{alg:ta}\eqref{alg:ta-adaptation} as
\begin{align*}
\log \eps_{k+1}
&=
\log \eps_k + \gamma_{k+1} H_{\eps_{k}}(\Theta_k, \Theta_k',Y_k')
\\ &=
\log \eps_k + \gamma_{k+1} h(\eps_k) + \gamma_{k+1} \eta_{k+1},
\end{align*}
where
$
H_\eps(\theta,\theta',y') \defeq \alpha^* - \alpha_\eps (\theta,\theta',y'),
$
with $\alpha_\epsilon$ defined in \eqref{eq:acc-epsilon}, with 'mean field'
$$
h(\eps)
\defeq
\int \pi^{(\eps)}(\ud \theta) q(\theta, \ud \theta') p^{(\theta')}(\ud y')  H_{\eps}(\theta,\theta', y'),
$$
and centred `noise' sequence
$
\eta_{k+1}
\defeq
H_{\eps_k}(\Theta_k, \Theta_k', Y_k') - h(\eps_k).
$
In this common framework for stochastic approximation algorithms, we can apply \citep[][Theorem 2.3]{andrieuMP}, which implies that the key lemma for the proof of convergence of the tolerance adaptive ABC-MCMC (Algorithm \ref{alg:ta}) essentially reduces to showing that the noise sequence $\eta_k$ is asymptotically controlled,
$$
\lim_{j\to \infty} \sup_{n\ge j} \bigg|\sum_{k=j}^n \gamma_k \eta_k\bigg| = 0,
\qquad \text{almost surely,}
$$
\citep[Lemma 20]{vihola-franks}.  This relies on various ancillary results, such as monotonicity of the map $\eps\mapsto h(\eps)$, continuity and contraction properties of the Markov kernels, and a generalisation of the `proposal augmentation' from Metropolis-Hastings chains \cite{schuster-klebanov,rudolf-sprungk18} to `proposal-rejection' chains.  Here, we call a kernel $K$ a `proposal-rejection' kernel if it is reversible  and can be written as
\begin{equation}\label{eq:proposal-rejection}
K(\theta, \ud \theta') = q(\ud \theta'|\theta) \alpha(\theta,\theta') +\Big(1-\int q(\ud \vartheta|\theta) \alpha(\theta,\vartheta)\Big) \delta_{\theta}(\ud \theta'),
\end{equation}
where $\alpha(\theta,\theta')\in [0,1]$ is a measurable function on $\T^2$.  By marginalising away the auxiliary variable in ABC-MCMC($\epsilon$) \eqref{alg:abc-mcmc}, we obtain such a `proposal-rejection' kernel, with
$$
\alpha(\theta,\theta') =
\min\bigg\{1, \frac{\pr(\theta') q(\theta| \theta')}{\pr(\theta) q(\theta'|\theta)}\bigg\}
L^{(1/\epsilon)}(\theta'),
$$
which is clearly not a Metropolis-Hastings kernel any longer.

Non-standard theoretical challenges of the tolerance adaptive ABC-MCMC (Algorithm \ref{alg:ta}) are that the invariant measure $\pi^{(1/\epsilon_k)}$ is changing at each iteration, and that the chain is technically a pseudo-marginal. Regarding this latter point, however, we do have simplification to independent refreshments of the auxiliary variable $y'$, because of the use of a simple cut-off function $\charfun{ d(\uarg,y^*)\le \epsilon}$ in the acceptance ratio.  
As mentioned in Section \ref{sec:abc-choice}, essentially, the convergence theorems for the adaptation are formulated in a simpler setting of uniform ergodicity, as well as for simultaneously geometrically ergodic `proposal-rejection' chains, obtained by only considering the marginal chain $(\Theta_k)_{k\ge 1}$ of the original chain $(\Theta_k, Y_k)_{k\ge 1}$, on possibly unbounded state space domains.  

\section{Discussion and directions for future work}\label{sec:discussion}
In this thesis, various old and new Monte Carlo estimators are presented.  A defining feature of the estimators suggested is that they involve an IS type correction of samples drawn according to an intermediate approximate distribution.  Basic convergence properties of the suggested estimators are established, and efficiency of these algorithms is studied and related to standard direct methods used hitherto commonly in practice.

There is still much interesting work that could be done in regards to the use of these estimators in different settings and with different approximations.  Experimental results  have been promising, and suggest further comparisons could be made,
for example, of 
PIMH and its IS analogue  \eqref{eq:ratio-estimator}.  In the parameter inference setting, there have been many MCMC implementations making use of an approximation by applying delayed acceptance \citep[see][Section 7.2]{franks-vihola}, but very few using MCMC-IS (see \cite{parpasUWT} for one other non-academic example).  One of the main goals of \cite{viholaHF} is to bring attention to MCMC-IS, that it represents a viable approach, which enjoys flexibility in implementation and theoretical backing.

In the filtering and smoothing context, the approach for optimal selection of Feynman-Kac model for the smoothing problem \cite{guarnieroJL} based on deterministic approximations, as used in \cite{viholaHF} and further developed in \cite{lindstenHV} for an extended class of models, could be further developed.  These approximations could also be based on various other non-linear filters and smoothers \citep[cf.][]{sarkka-filtering-smoothing}.

There are various directly applicable innovations which could be incorporated into MCMC-IS, and we mention a few.
Quasi-Monte Carlo may be helpful in MCMC-IS, whether in the MCMC \citep[cf.][]{schwedes-calderhead} or in the PF \cite{gerber-chopin}.
Work on exact simulation \cite{beskos-roberts} techniques for diffusions \cite{beskosPR,beskosPRF,fearnheadLRS} (see also the recent preprint \cite{wang-rao-teh}) and jump-diffusions \cite{goncalvesLR} using continuous-time IS techniques is showing progress, and suggests parameter inference methods for partially observed versions could be developed, at least in the one-dimensional setting, using the MCMC-IS framework, with IS correction based on a PF using exact simulation dynamics, or based on other types of randomised weights, which may freely assume negative values in the IS correction.

It would be of interest to adapt the tuning guidelines \cite{doucetPDK} (see also \cite{sherlockTRR}) for the PF when used in the PMMH, to the case when used within MCMC-IS.  The formula \eqref{eq:is-asvar} for the MCMC-IS asymptotic variance, which decomposes into marginal chain and IS correction parts, could also be useful in this regard.       
More generally, beyond PMMH, it would be beneficial to use better scaling MCMCs within MCMC-IS, for example, particle Gibbs \cite{andrieuDH}, which is known to scale very well with backward sampling \citep[cf.][]{leeSV}.  Additional annealing steps may be useful, as part of the Metropolis-Hastings with asymmetric acceptance ratio (MHAAR) approach \citep[cf.][]{andrieuDYC}

In \cite{franks-vihola},
more practical examples could be given showing bounds of likelihood ratios and usefulness of the results in practice.
Further comparisons could be made, for example, with annealed IS \cite{neal-annealed} correction versus multi-stage DA \cite{banterle-grazian-lee-robert}. Two different extensions of traditional DA correction were introduced in \cite{franks-vihola}, and it would be interesting to study the stability properties of these new DA corrections, for example,  along the lines of \cite{andrieu-roberts,sherlock-lee}.
Other more sophisticated reversible chains as in \cite{andrieuDYC} with IS correction could be considered and compared.  The effect of debiasing tricks  \cite{glynn-rhee} could be compared between  MCMC-IS and pseudo-marginal type MCMC, where the coupling time integral to the debiasing approach may be considerably less for MCMC-IS if Phase \ref{alg:is-one} is based on deterministic approximation and Phase \ref{alg:is-two} involves noisy unbiased estimators.

There are many settings where there is a multilevel type structure and the debiasing techniques can be applied.  In the joint inference setting, the IS-debiasing method as presented in \cite{franksJLV} allows for an efficient debiasing strategy for joint inference using Euler approximations.
The results could be generalised to It\^{o} diffusions with time-dependent path-dependent coefficients, and to general resampling schemes in the $\Delta$PF besides multinomial resampling.
It would be nice to apply the IS-debiasing strategy in various settings, for example, to jump-diffusions \citep[cf.][]{dereich-gaussian,jasraLO}.  The coupling \cite{jasraKLZ} and multilevel approach to the (unnormalised) smoothing problem \cite{franksJLV}, with possible randomised MLMC correction \cite{mcleish,rhee-glynn}, could be applied, for example, to the problem of calculation of normalisation constants \citep[cf.][]{moralJLZ} important for model selection.
The optimal choice of coupled dynamics and potential could be studied, where we remark that the coupled potentials may be made level dependent, which is an additional degree of freedom.  It would also be of interest to study stability and limit theorems \citep[][]{chopin-central,del-moral,douc-moulines} of these coupled PFs \citep[cf.][]{jasra-yu} based on change of reference measure and IS reweighting for use in unnormalised multilevel estimators as in \cite{franksJLV}.  

We are currently looking into  optimal tuning of the regularisation constant in the approximate likelihood estimator within MCMC-IS, which is connected to `defensive importance sampling' \cite{hesterberg}. 
The question of efficiency and proper allocation of resources of the MCMC-IS carries over to the multilevel and PF setting, where additionally multilevel aspects play a r\^{o}le.  The question of optimal scaling particles versus level in the sub-canonical regime associated to Euler approximations was not entirely conclusive in \cite{franksJLV}.  It would be interesting to study this phenomenon in more depth.  This may entail adapting the non-canonical CLT of \cite{zhengBG} in the diffusion setting to the partially observed diffusion setting where number of particles and particle approximation variances are additional factors.

Applied in the ABC context in \cite{wegmannLE}, the post-correction (or trimming) over a range of tolerances is a methodological approach applicable in other Monte Carlo settings where IS can be applied at small additional cost, for example, in the MLMC context, with
the sum of multilevel increments computed sequentially
over an increasing range of the fine tolerances, with corresponding plots.
In such settings, it may also be possible to derive analogous approximate confidence intervals for the resulting estimators as in \cite{vihola-franks}.
The tolerance adaptive ABC-MCMC in \cite{vihola-franks} was based on targeting a user-specified overall acceptance probability, and we chose a target close to the rule of thumb from the more general random walk PM literature \cite{sherlockTRR}.  It may be interesting to adapt the assumptions of \cite{doucetPDK,sherlockTRR} to ones more resembling the ABC context, in order to find a perhaps different `rule of thumb' for ABC-MCMC.
The tolerance adaptation was also found to benefit the covariance adaptation during the burn-in, likely due to the improved mixing in the initial stages of the algorithm.  It would be of interest to study this phenomenon and the interplay of different optimisation criteria in more depth, following, for example, the theoretical developments of adaptive MCMC as in \cite{andrieu-moulines,andrieuMP,andrieu-thoms}.

The `proposal-rejection' chains \eqref{eq:proposal-rejection}, which were considered for DA correction \cite{franks-vihola} and  `proposal augmentation' \cite{vihola-franks}, are generalisations of Metropolis-Hastings chains, which include DA \citep{banterle-grazian-lee-robert}, PM \cite{andrieu-roberts}, MHAAR \cite{andrieuDYC} and marginalised ABC-MCMC \cite{vihola-franks}.  Although `proposal-rejection' chains technically include pseudo-marginal chains, we lay here particular emphasis on the possible course of study of (simpler) chains on the marginal (parameter) space, without auxiliary variable extensions like in pseudo-marginal MCMC.
Many results worked out for Metropolis-Hastings can likely be extended to the marginal-space `proposal-rejection' setting.  For example, the waste-recyclers of \cite{delmas-jourdain,rudolf-sprungk18,schuster-klebanov}, originally for Metropolis-Hastings, could be extended to `proposal-rejection' chains.
Some convergence analysis has been done for pseudo-marginal Metropolis-Hastings chains \citep[cf.][]{andrieu-roberts,andrieu-vihola-convergence} and some of this type of analysis could possibly be adapted to marginal-space `proposal-rejection' chains.
Following the line of argument  of \cite{jarner-hansen,roberts-tweedie},
who show geometric ergodicity of symmetric random walk Metropolis-Hastings essentially if the target has exponential or lighter tails and a certain contour condition holds, it would be interesting to work out conditions for a similar type of result for the more general sub-class of marginal-space `proposal-rejection' chains.


\section{Summary of articles}\label{sec:summary}
\subsection[sec]{\for{toc}{Article [A]: Importance sampling type estimators...}\except{toc}{Article [A]}}

Convergence properties are established for Markov chain Monte Carlo (MCMC) algorithms using an additional importance sampling (IS) type correction of approximate sample output of the Markov chain.
Included is the interesting case where the approximate chain itself stems from a pseudo-marginal chain.
The asymptotic variance from the proven central limit theorems is shown to decouple over the approximate marginal chain and the IS correction, which can be useful for questions of optimal allocation of computational resources.
Particular strengths of the approach are highlighted, such as the efficient use of a jump chain, thinning, and straightforward parallelisation.  Abstract properties of the augmented Markov chains corresponding to the MCMC-IS method are established.  Experiments in state space models compare the MCMC-IS method with existing popular direct methods, and show the viability of the MCMC-IS approach in the state space models context.

\subsection[sec]{\for{toc}{Article [B]: Importance sampling correction versus...}\except{toc}{Article [B]}}
The asymptotic variance of the MCMC-IS is compared to that of the direct MCMC methods.  This is based on an extension of the existing covariance comparison result for direct chains to the context of comparison of one MCMC-IS to one direct chain.  The extension also allows for use of unbiased estimators in the MCMC-IS Phase 1 and 2, as well as the use of a jump chain.  Provided examples show that there can be no strict ordering between MCMC-IS and direct MCMC, as either may perform arbitrarily better than the other.  Theoretical results are provided, which show upper and lower bounds for the MCMC-IS asymptotic approach in relation to an analogous direct MCMC method.  The upper bound is satisfied in practice when approximations are reasonably accurate, and provides guarantees for the MCMC-IS asymptotic variance in terms of direct pseudo-marginal and delayed acceptance analogues.  In the latent variable setting, this is the case in the sense of  finite supremum norm of the ratio of likelihoods.  Ergodicity and mixing of the MCMC-IS is shown to be less affected by noise of the Phase 2 unbiased estimators compared to pseudomarginal direct MCMC.
The results help justify the viability of the MCMC-IS approach as a competing method to a direct approach.

\subsection[sec]{\for{toc}{Article [C]: Unbiased inference for hidden Markov model...}\except{toc}{Article [C]}}
The question of joint inference for a challenging class of state space models is considered, where the underlying process is a diffusion process arising as a solution to a stochastic differential equation, which can not be simulated exactly.  Noisy non-linear observations are obtained at some discrete points in time.  Bayesian inference is performed using the IS debiasing approach, where, namely, an IS type correction, based on debiased multilevel Monte Carlo, a particle filter coupling, and Euler approximations, is used for an approximate MCMC targeting a coarse-model approximate distribution.  Convergence of the method to the exact posterior is verified under standard conditions on the state space model and Euler type approximations found in the literature.  From asymptotic efficiency and cost considerations, suggested allocations for computational resources are given, which help ensure efficient use of the algorithm.     


\subsection[sec]{\for{toc}{Article [D]: On the use of ABC-MCMC...}\except{toc}{Article [D]}}
The use of a slightly inflated tolerance is suggested in the context of approximate Bayesian computation (ABC) MCMC, along with subsequent post-correction  based on trimming or IS correction of the sample output, over a (continuous) range of decreasing tolerances.  Approximate confidence intervals for the resulting estimators are provided, which enjoy theoretical backing as well as good coverage in the experiments considered.  An adaptive ABC-MCMC is also proposed, which finds a suitable (inflated) tolerance based on acceptance rate as the proxy.   Convergence theorems for the adaptation under simple and more general conditions are provided.  The tolerance adaptation worked well when used together with proposal covariance adaptation, in experiments which confirmed the suitability of the method based on adaptive ABC-MCMC and post-correction.

\newpage
\bigskip
\normalsize
\bibliographystyle{abbrv}

\end{document}